\documentclass[%
 reprint,
superscriptaddress,
 amsmath,amssymb,
 aps,
]{revtex4-1}

\usepackage{graphicx}% Include figure files
\usepackage{dcolumn}% Align table columns on decimal point
\usepackage{bm}
\usepackage{physics}% bold math
\usepackage{hyperref}% add hypertext capabilities
\usepackage{siunitx}
\usepackage[caption=false]{subfig}
\begin{document}

\preprint{APS/123-QED}

\title{Competitive effects between gravitational radiation and mass variation for two-body systems in circular orbits}% Force line breaks with \\

\author{Baptiste Blachier}
\email{baptiste.blachier@ens-lyon.fr}
 \affiliation{Laboratoire de Physique Subatomique et de Cosmologie, Univ. Grenoble-Alpes, CNRS/IN2P3 \\ 53, avenue des Martyrs, 38026 Grenoble cedex, France}
 \affiliation{Department of Physics, \'Ecole Normale Supérieure de Lyon, 69364 Lyon, France}
\author{Aurélien Barrau}%
 \email{barrau@lpsc.in2p3.fr}
\author{Killian Martineau}
\email{martineau@lpsc.in2p3.fr}
\author{Cyril Renevey}
\email{renevey@lpsc.in2p3.fr}
\affiliation{Laboratoire de Physique Subatomique et de Cosmologie, Univ. Grenoble-Alpes, CNRS/IN2P3 \\ 53, avenue des Martyrs, 38026 Grenoble cedex, France}

\date{\today}% It is always \today, today,

\begin{abstract}
This work provides, at lower order, general analytical solutions for the orbital separation, merging time, and orbital frequency of binary systems emitting gravitational waves while being submitted to mass variations. Specific features, depending on the exponent of the mass derivative, are investigated in details. Two phenomenologically interesting cases are explicitly considered : i) binaries formed by two light primordial black holes submitted to Hawking evaporation and ii) bodies driven by a Bondi accretion of phantom dark energy. It is shown that three different regimes arise, including an intricate non-monotonic behaviour of the system. We study subtle imprints that could be associated with those phenomena. A careful analysis of the conditions of validity of the different hypotheses performed is finally carried out.
\end{abstract}

\maketitle

\section{\label{sec:intro}Introduction}
Newtonian orbits are stable for objects with constant masses. This is obviously not the case anymore if the considered masses become time-dependent, whatever the reason for this variation. This also becomes wrong in General Relativity (GR) when the system radiates energy through gravitational waves. The very existence of this energy has been intensively debated, even quite recently \cite{Duerr:2019fsv}, as one might wonder if objects following geodesics -- thus being ``force-free" -- can dissipate energy. It is however well settled since the works performed by Penrose, Bondi, and Sachs in the 1960s, although the conceptual arguments are somehow subtle and the technicalities quite involved \cite{Gomes:2023xda}. This work aims at investigating -- at lowest order -- the detailed evolution of the orbits of systems submitted both to the emission of gravitational waves (GWs) and to mass variations.

One of the most striking sources of GWs is the coalescence of black holes (BHs), which has been exhaustively studied theoretically and has proved to be experimentally fruitful since most of the GW signals detected so far by LIGO and Virgo collaborations come from this phenomenon (see, {\it e.g.}, \cite{LIGOScientific:2020ibl,LIGOScientific:2021djp}). Although the most simple and straightforward cases are those of binaries with constant masses, inspiral binaries with time-varying masses and their consequences on the emitted gravitational radiation have also been studied in the literature \cite{Holgado:2019ndl, Barausse:2014tra, Mersini-Houghton:2008cul, Enander_2010, ONeill:2008sat, Macedo:2013qea, Sarkar:2019tdf, Sarkar:2022jtn, Yagi:2011yu, Simonetti:2010mk, Yuan:2021xdi}. It is well-known that the emission of GWs tend to make the two bodies of the binary come closer to each other, until they ultimately coalesce \cite{Maggiore:2007ulw}. However, the time-varying mass can have the converse effect, depending on whether the bodies gain or lose mass. If they gain mass, for instance through matter accretion, they inspiral, thus enhancing the effect of the emission of GWs. On the other hand, if they lose mass, they outspiral: in that case, mass loss produces an antagonistic effect to the one of gravitational radiation. 

In this note, we focus on presenting a comprehensive view of the dynamics induced by the competitive effects of gravitational radiation and mass loss, using Newtonian analysis and treating the binary system as Keplerian, as well as the characteristic features of the imprint it produces on the emitted GWs. In section \ref{sec:sec2}, 
%notably inspired by \cite{Enander_2010}, 
we review how one can theoretically deal with the coupling between mass variation and the backreaction produced by the emission of gravitational waves. However, unlike previous works -- see Refs. \cite{Mersini-Houghton:2008cul, Enander_2010, ONeill:2008sat, Macedo:2013qea, Sarkar:2019tdf, Sarkar:2022jtn} for accreting bodies or Refs. \cite{Yagi:2011yu, Simonetti:2010mk, Yuan:2021xdi} for evaporating BHs in the context of braneworld models and extra-dimensions scenarios -- we present a generic analytic solution of the differential equation satisfied by the orbital separation when the mass variation is taken into account. This leads in turn to analytic solutions for the frequency and strain of the emitted GWs.

For cosmology and phenomenology, we focus specifically on two situations that are driven by such a competition. In section \ref{sec:sec3}, we investigate the case of inspiral binaries of light primordial black holes (PBHs) submitted to Hawking evaporation. Due to their low mass, they are good candidates as sources of high-frequency GWs \cite{Aggarwal:2020olq} -- in addition to standing as plausible dark matter candidates and valuable probes of the early universe.

The second case of interest is the one of binaries of BHs accreting phantom energy ({\it i.e.} violating energy dominance conditions) which makes their mass \textit{decrease} \cite{Babichev:2004yx}. Although the accretion of a fluid onto a black hole is a long-standing problem in astrophysics \cite{Bondi1952}, the study of this so-called phantom energy has proved particularly relevant in the context of dark energy \cite{Babichev:2005py}. Assuming that the latter can be described by a perfect fluid of density $\rho$ and pressure $p$, an equation of state parameter $w = p/\rho$ smaller than -1 would make dark energy behave as phantom energy: accretion would decrease the mass. Coining down the value of $w$ is a hot spot in modern cosmology \cite{Carroll:2003st} and doing so by examining small-scale systems like BHs binaries was proposed in \cite{Mersini-Houghton:2008cul, He:2009jd} and questioned in \cite{Enander_2010}. In section \ref{sec:sec4}, we refine the analysis and draw a clear conclusion.

Beyond the specific examples chosen for their physical relevance, this study aims at offering an exhaustive classification of possible behaviors of binary systems emitting gravitational waves and submitted to mass variation effects (either accretion or radiation) described by a (positive or negative) power law. These generic results are presented in section \ref{sec:sec5}

Eventually in section \ref{sec:sec6}, we show how accounting for a time-varying mass changes the formal expression of the radiated gravitational power and in section \ref{sec:sec7} we discuss how the results of section \ref{sec:sec5} are changed when considering non-identical masses presenting a high hierarchy.

\section{\label{sec:sec2}Coupling mass variation and GWs emission}
The two-body problem with variable mass, for point-like entities, was extensively studied in Ref. \cite{HADJIDEMETRIOU196634, Verhulst1975}. When the variation of mass is isotropic, the equation of motion is given by
\begin{equation}
    \dv[2]{\bm{r}}{t} = - \frac{Gm_{\mathrm{tot}}(t)}{R^{3}}\bm{r},
\end{equation}
where $\bm{r} = \bm{r}_{2}– \bm{r}_{1}$, thus $R \equiv \abs{\bm{r}}$ is the orbital separation between the two bodies of respective mass $m_1$ and $m_2$, the total mass being $m_{\mathrm{tot}}(t) \equiv m_{1}(t) + m_{2}(t)$. For orbital separations much larger than the Schwarzschild radii of the black holes, $R \gg R_{\mathrm{s}} = 2Gm/c^{2}$, we can describe the binary system using Keplerian dynamics. If $\mu \equiv m_{1} m_{2} / (m_{1} + m_{2})$ stands for the reduced mass, the orbital angular momentum of the system, for a circular orbit, whose generic expression is given by 
\begin{equation}
    J_{\mathrm{orb}} = \mu \sqrt{Gm_{\mathrm{tot}} R},
    \label{eq:K}
\end{equation}
is conserved, even in the case of variable masses \cite{HADJIDEMETRIOU196634}. If we consider that the two bodies have the same mass $m(t)$, then the conservation of the orbital angular momentum $\dot{J}_{\mathrm{orb}} = 0$ provides 
\begin{equation}
    \dot{R} = - 3\frac{\dot{m}}{m} R,
    \label{eq:dotR}
\end{equation}
showing that in the case of mass loss ($\dot{m} <0$) the two bodies outspiral, {\it i.e.} they drive away from each other, which is consistent with Refs. \cite{HADJIDEMETRIOU196634, McWilliams:2009ym, Simonetti:2010mk, Yuan:2021xdi}. From now on, we shall always assume, unless otherwise stated, that the two bodies have same mass $m(t)$. This is the most interesting situation and it is phenomenologically sound -- at least as a first approximation -- if the mass spectrum of PBHs is nearly monochromatic (see \cite{Carr:2020gox} for a review of formation mechanisms). Only in section \ref{sec:sec7}, will we drop this assumption in order to examine how certain results presented in the following are modified when the binary system is made of bodies of different masses, but presenting a strong hierarchy (typically one mass dominates over the other).

The mass loss, in addition to separating the two bodies also induces a change in the orbital energy $E_{\mathrm{orb}} = - Gm^{2}/(2R)$, which is the total energy of the system, sum of kinetic and potential energy of the orbit. For variable masses, the variation of orbital energy $-\dv{E_{\mathrm{orb}}}{t}$ must be equal to the power associated to mass loss $P_{\mathrm{ml}}$, thus leading to
\begin{equation}
    P_{\mathrm{ml}} = \frac{5}{2} \frac{G\dot{m}m}{R}.
\label{eq:P_ml}
\end{equation}
On the other hand, the emission of GWs costs energy which is taken from the orbital energy of the system, carried away at a rate \cite{Maggiore:2007ulw}
\begin{equation}
    P_{\mathrm{gw}}(t) = \frac{64}{5} \frac{G^{4}}{c^{5}} \frac{m^{5}(t)}{R^{5}}.
\label{eq:Pgw}
\end{equation}
In fact, $m$ having an explicit time-dependence, it affects the form of the power radiated by GWs thus modifying Eq. \eqref{eq:Pgw}, as already noted in \cite{Enander_2010}, through several corrective terms. We discuss this point in further detail in section \ref{sec:sec5}, and show how the choice of the form \eqref{eq:Pgw} for $P_{\mathrm{gw}}$ is connected to the condition of circularity of the orbit.

To consider the concomitant effect of gravitational radiation and mass variation, one can use the conservation of energy and write \cite{Enander_2010}
\begin{equation}
     - \dv{E_{\mathrm{orbit}}}{t} = P_{\mathrm{ml}} + P_{\mathrm{gw}},
\end{equation}
with $P_{\mathrm{ml}}$ and $P_{\mathrm{gw}}$ respectively given by Eqs. \eqref{eq:P_ml} and \eqref{eq:Pgw}. Taking the derivative of the orbital energy leads to
\begin{equation}
    \dot{R} = -\frac{128}{5} \frac{G^{3}}{c^{5}} \frac{m^{3}}{R^{3}} - 3\frac{\dot{m}}{m}R,
\label{eq:ED_general}
\end{equation}
which is the general differential equation relevant for our problem. 

This agrees with the result of Ref. \cite{Enander_2010} -- up to the corrected prefactor 3 in the last term. This deserves a brief specific discussion. Equation \eqref{eq:dotR} holds only if the masses are identical. There are two other limit cases allowing for simple formulas (as can immediately be seen from the expression of the angular momentum). They correspond to a strong mass hierarchy with asymmetrical losses. We specifically investigate those cases in the last section of this work. We will show there that if the varying mass is the small one, the prefactor 2, inappropriately used in \cite{Enander_2010} indeed appears. Hence the probable origin of the mistake.
%Let us call $m_1$ the stable mass and $m_2$ the varying one. If $m_1\gg m_2$, the evolution becomes $\dot{R}=-2(\dot{m_2}/m_2)R$ whereas is reads $\dot{R}=-(\dot{m_2}/m_2)R$ if $m_1\ll m_2$. This is probably the origin of the mistake in \cite{Enander_2010}. 

If the functions $m(t)$ and $\dot{m}(t)$ are explicitly known, this differential equation, although non-linear, can be integrated since it is a Bernoulli differential equation of the form $y' + P(t) y = Q(t) y^{n} $ with $n=-3$ whose general solution is known. In the following two sections, we shall focus 
on mass losses described by rates of the form $\dot{m} \propto -1/m^{2}$ (see section \ref{sec:sec3})  and $\dot{m} \propto -m^{2}$ (see section \ref{sec:sec4}). More generic results shall be found in section \ref{sec:sec5}.

Throughout all this work, we call ``merging" or ``coalescence" the situation corresponding to the precise vanishing of the orbital separation ($R=0$), and not to the ``contact" of horizons ($R=4Gm/c^2$). For most observables this makes nearly no difference at all, but it is important when considering the end of the process.

\section{\label{sec:sec3}Inspiral binaries with evaporating BHs}
Because of their small masses, primordial black holes can be particularly sensitive to the Hawking evaporation process (see \cite{MacGibbon_PBH, Halzen:1991uw} for pioneering works and \cite{Carr:2020gox} for a recent review). Quantum field theory in curved spacetimes predicts that black holes evaporate with a temperature \cite{Hawking:1975vcx, Birrell:1982ix}
\begin{equation}
	T = \frac{\hbar c^{3}}{8\pi G k_{\mathrm{B}} m},
\end{equation}
for a BH of mass $m$, $\hbar$ being the Planck constant, $k_{\mathrm{B}}$ the Boltzmann constant, $c$ the speed of light, and $G$ the gravitational constant. Even though for typical solar-mass BHs, Hawking evaporation is too weak to play any significant role in the dynamics of binaries, for two-body systems made of PBHs, it could produce non-negligible effects on the overall dynamics, and leave a possible imprint on the emitted gravitational waves. The Hawking process leads to a mass loss at the rate \cite{MacGibbon_PBH, Halzen:1991uw}
\begin{equation}
    \dot{m} = - \frac{\alpha_{\mathrm{H}}}{m^{2}},
\end{equation}
where $\alpha_{\mathrm{H}}$ accounts for the degrees of freedom of each particle contributing to the evaporation. For simplicity, we shall assume that $\alpha_{\mathrm{H}}$ is a constant, which is a good approximation. In this case, the differential equation is separable and integrates into 
\begin{equation}
    m(t) = m_{0} \left(1 - \frac{t}{t_{\mathrm{ev}}}\right)^{1/3},
\label{eq:hawking_mass}
\end{equation}
where $m_{0}$ represents the initial mass of the BH whereas $t_{\mathrm{ev}} \equiv m_{0}^{3}/(3\alpha_{\mathrm{H}})$ corresponds to the time of evaporation, {\it i.e.} the typical time it takes for the BH to evaporate completely. The other time scale of the problem under scrutiny is the \textit{time of coalescence} which accounts for the limited life duration of the binary system. However, contrary to $t_{\mathrm{ev}}$ which only depends on constants fixed during the initial setup (namely the initial mass $m_{0}$), the time of coalescence has an explicit dependence on the mass $m$ (and not only on the initial mass $m_{0}$). Otherwise stated, if $t_{\mathrm{ev}}$ is not to be altered by the dynamics of the system, the time of coalescence will necessarily be affected by the mass loss. Consequently, we will call $t_{\mathrm{cc}}$ the time of coalescence of the binary system \textit{if} the two BHs were of constant mass $m_{0}$, with an initial orbital separation $R_{0}$, {\it i.e.} \cite{Maggiore:2007ulw}
\begin{equation}
    t_{\mathrm{cc}} = \frac{5}{512} \frac{c^{5} R_{0}^{4}}{G^{3}m_{0}^{3}},
\label{eq:def_tcc}
\end{equation}
while $t_{\mathrm{coal}}$ will denote the real time of coalescence, that is to say the one taking into account the mass loss, which is likely to be different from $t_{\mathrm{cc}}$ and yet to be determined.

\subsection{Evolution of the orbital separation}
Specifying \eqref{eq:ED_general} to the case of Hawking evaporation provides
\begin{equation}
    \dot{R} = - \frac{128}{5} \frac{G^{3}}{c^{5}} \frac{m^{3}}{R^{3}} + \frac{3\alpha_{\mathrm{H}}}{m^{3}}R.
\label{eq:ED_R_2}
\end{equation}
Evaluating the above at an initial time $t_{0} = 0$ gives
\begin{equation}
    \frac{\dot{R}}{R} \bigg\rvert_{t_{0} = 0} = - \frac{1}{4t_{\mathrm{cc}}} + \frac{1}{t_{\mathrm{ev}}}.
\label{eq:ED_R_3}
\end{equation}
If one assumes that the system is prepared -- which boils down to giving ourselves $m_{0}$ and an initial separation $R_{0}$ -- such that $t_{\mathrm{ev}} < 4t_{\mathrm{cc}}$, then the evaporation process initially dominates. Since the evaporation increases $R$ and since  the GW part is in $m^{3}/R^{3}$ -- see Eq. \eqref{eq:ED_R_2} -- when $m$ diminishes and $R$ increases, both these variables contribute to the dwindling of this term, that will thus never be able to grow again to counterbalance the outspiraling effect. In conclusion, when $t_{\mathrm{ev}} < 4t_{\mathrm{cc}}$, the system initially outspirals and continues to do so until the PBHs eventually evaporate completely, leading to the disappearance of the binary.

In the other case, no definite conclusion can be stated by the sole inspection of Eq. \eqref{eq:ED_R_2}. When $4t_{\mathrm{cc}} < t_{\mathrm{ev}}$, we are in a regime where, initially, the radiation of GWs dominates, thus $R$ decreases. But as $R$ and $m$ diminish, there might be a counterbalancing effect due to the term of relative mass loss. Using the explicit expression \eqref{eq:hawking_mass} and solving the Bernoulli differential equation provides
\begin{equation}
    R(t) = R_{0} \left(\frac{t_{\mathrm{ev}}}{t_{\mathrm{ev}}-t}\right) \left(1 + \frac{1}{6} \frac{t_{\mathrm{ev}}}{t_{\mathrm{cc}}} \left[\left(1- \frac{t}{t_{\mathrm{ev}}}\right)^{6} -1 \right] \right)^{1/4}.
\label{eq:R_analytic}
\end{equation}
When $t_{\mathrm{cc}} \to \infty$, the second bracket boils down to unity and we recover the solution where no gravitational waves are emitted with the sole effect of mass loss leading to an outspiralling dynamics. Equation (\ref{eq:R_analytic}) is the exact analytical solution to the problem under consideration.

\begin{figure}[h]
    \centering
    \includegraphics[width=0.48 \textwidth]{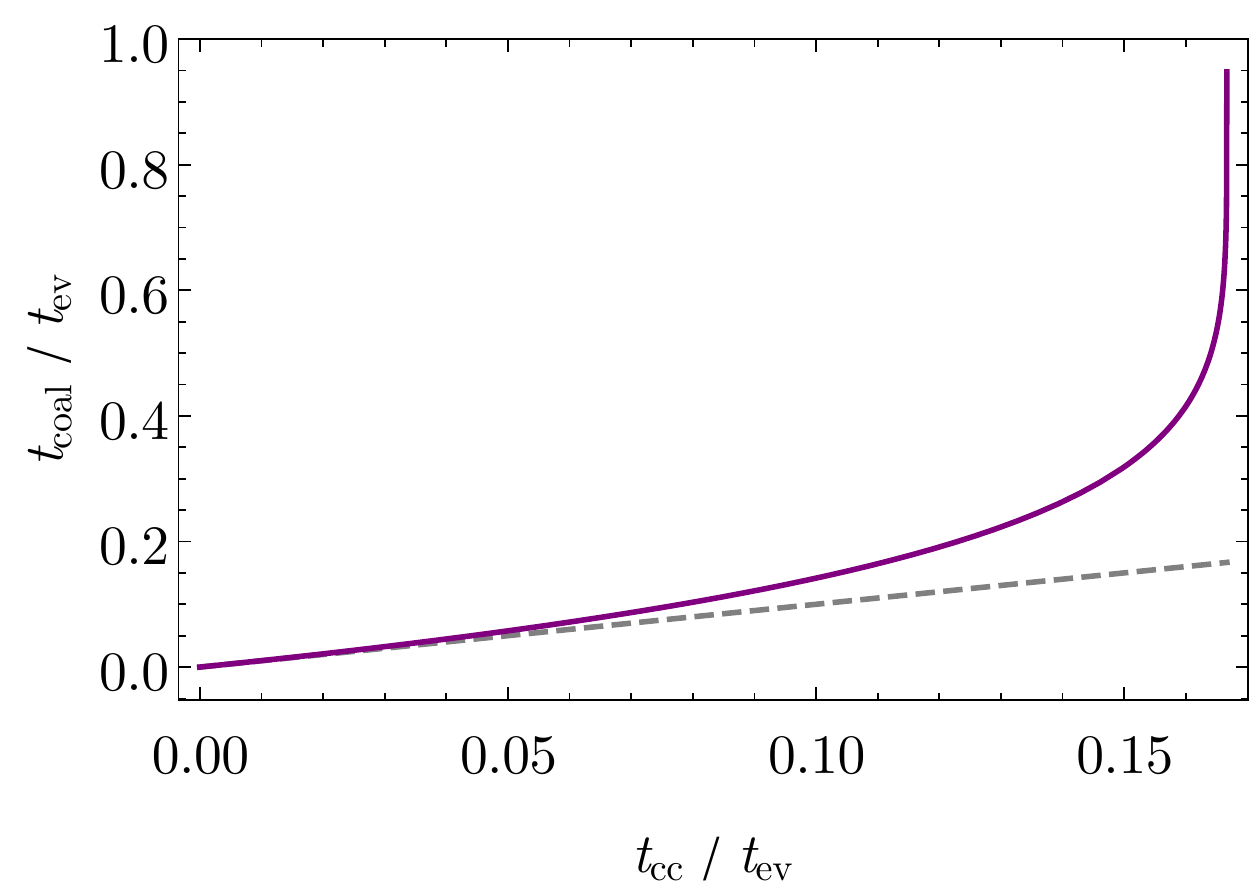}
    \caption{Effective time of coalescence $t_{\mathrm{coal}}$ as given by Eq. \eqref{eq:tau_effective_coal} divided by the time of evaporation $t_{\mathrm{ev}}$ (purple curve) in function of the ratio $t_{\mathrm{cc}}/t_{\mathrm{ev}}$. At $t_{\mathrm{cc}}/t_{\mathrm{ev}} = 1/6$, this function is finite equal to 1, and for higher values, it ceases to be mathematically well-defined. The grey dashed line is the identity function.}
    \label{fig:t_coal}
\end{figure}

The equation $R(t) = 0$ has for solution 
\begin{equation}
    t_{\mathrm{coal}} = t_{\mathrm{ev}} \left( 1 - \left[1 - 6\frac{t_{\mathrm{cc}}}{t_{\mathrm{ev}}} \right]^{1/6} \right),
\label{eq:tau_effective_coal}
\end{equation}
which corresponds to the real time of coalescence. Mathematically, this solution is well-defined only if $1-6t_{\mathrm{cc}}/t_{\mathrm{ev}} > 0$ which leads to the condition 
\begin{equation}
    t_{\mathrm{ev}} > 6t_{\mathrm{cc}}.
\label{eq:condition}
\end{equation}
If Eq. \eqref{eq:condition} is satisfied, the equation $R(t) = 0$ always admits a solution and thus the two PBHs do coalesce, in a similar fashion to what is observed in the case of constant masses. However, as shown in Fig. \ref{fig:t_coal}, because of the mass loss which renders the system less tightly bound, $t_{\mathrm{coal}}$ is always larger than $t_{\mathrm{cc}}$.

This immediately raises the question of what happens in the small window of parameters for which $4t_{\mathrm{cc}} < t_{\mathrm{ev}} < 6t_{\mathrm{cc}} $. Equation \eqref{eq:ED_R_3} indicates that we are then in a regime where the emission of gravitational waves \textit{initially} dominates, thus leading, in a first step, to the decrease of the orbital separation. However, since the condition \eqref{eq:condition} is not fulfilled, the equation $R(t) = 0$ does not admit any solution, which indicates that the system will never merge. This is due to the evaporation term that gets progressively more important and eventually dominates the GW emission. Once the evaporation process leads, there is no coming back and it ultimately makes the two PBHs outspiral, until they completely disappear. This is an interesting and highly non-trivial case where the orbital separation evolution is non-monotonic.

\begin{figure}[h]
\centering
\includegraphics[width=0.48 \textwidth]{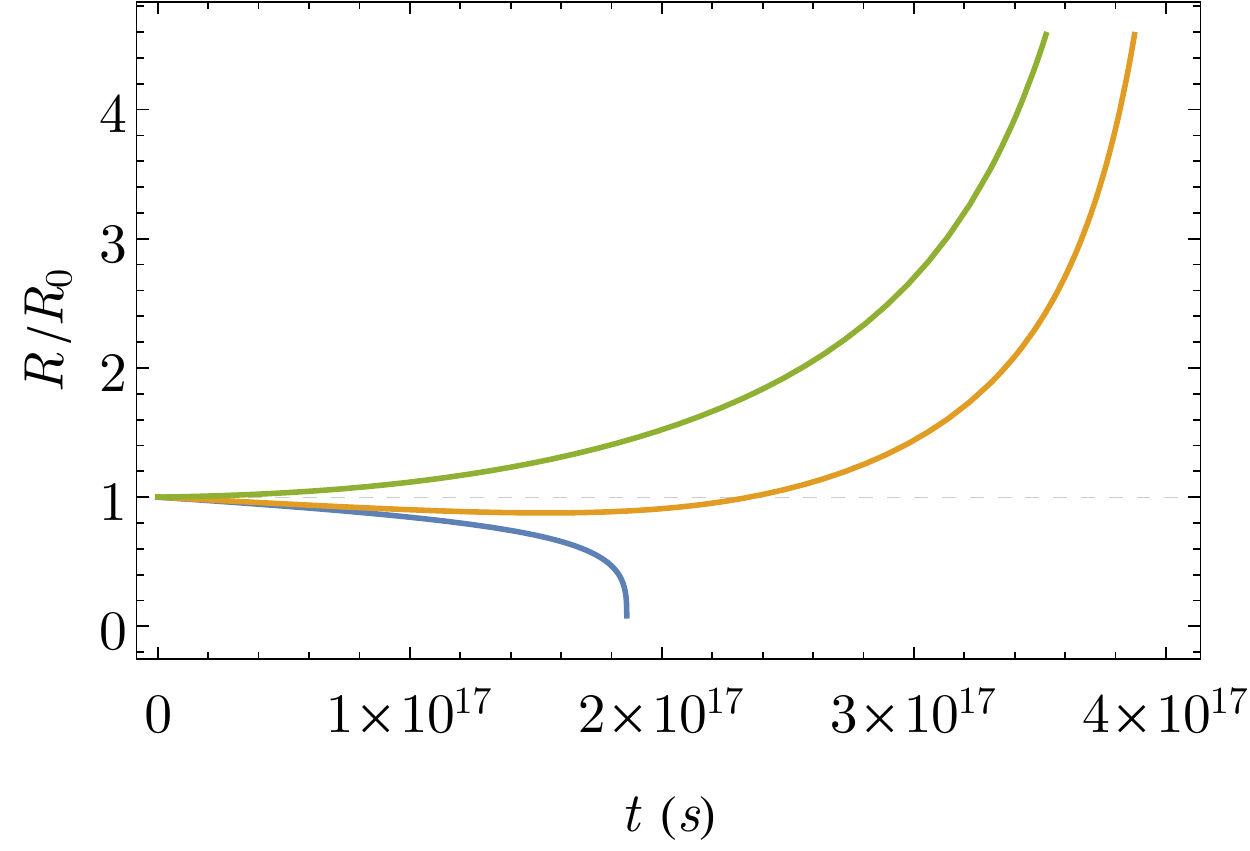}
\caption{\label{fig:R_equal_mass} Evolution of the orbital separation $R$ (normalized to the initial orbital separation $R_{0}$) between two evaporating black holes with initial mass $m_{0} = \SI{e12}{kg}$, as a  function of time (in seconds). Three distinct regimes arise : for $t_{\mathrm{ev}} < 4t_{\mathrm{cc}}$ (green curve) the binary outspirals whereas for $t_{\mathrm{ev}} > 6t_{\mathrm{cc}} $ (blue curve) it inspirals following a trend similar to the standard case of constant masses. For $4t_{\mathrm{cc}} < t_{\mathrm{ev}} < t_{\mathrm{cc}} $ (orange curve), one has an intermediate regime mixing the two behaviours, with an unusual non-monotonic evolution.}
\end{figure}

To sum up, there are three distinct regimes (see Fig. \ref{fig:R_equal_mass}) :
\begin{itemize}
    \item If $t_{\mathrm{ev}} < 4t_{\mathrm{cc}}$, the system outspirals due to the domination of the evaporation process and $R$ increases with a final divergence corresponding to the full evaporation of the two PBHs;
    \item If $4t_{\mathrm{cc}} < t_{\mathrm{ev}} < 6t_{\mathrm{cc}} $, the two PBHs come closer together in a first step due to the emission of GWs and then, as the evaporation takes the lead back, they ultimately outspiral as in the first case; and
    \item If $t_{\mathrm{ev}} > 6t_{\mathrm{cc}} $ the emission of GWs dominates entirely, leading to the merger of the system, in a similar fashion to what is observed in the case of constant masses. The only real influence of the mass loss is to increase the time of coalescence.
\end{itemize}

The above inequalities can be conveniently expressed as conditions on the initial orbital separation $R_{0}$ for a given initial mass $m_{0}$ by using Eq. \eqref{eq:def_tcc}. If one defines 
\begin{equation}
    R_{1} \equiv \left(\frac{256}{45} \frac{G^{3}}{c^{5} \alpha_{\mathrm{H}}}\right)^{1/4} m_{0}^{3/2},
\end{equation}
and
\begin{equation}
    R_{2} \equiv \left(\frac{128}{15} \frac{G^{3}}{c^{5} \alpha_{\mathrm{H}}}\right)^{1/4} m_{0}^{3/2},
\end{equation}
the ouspiralling regime is reached when $R_{0} > R_{2}$, the inspiralling one when $R_{0} < R_{1}$, while the intermediate regime corresponds to $R_{1} < R_{0} < R_{2}$. This might be relevant when evaluating the merging rate of PBHs \cite{Kocsis:2017yty,Raidal:2018bbj,Gow:2019pok}.

\subsection{Analysis of the frequency}
\begin{figure}[h]
\centering
\includegraphics[width=0.48 \textwidth]{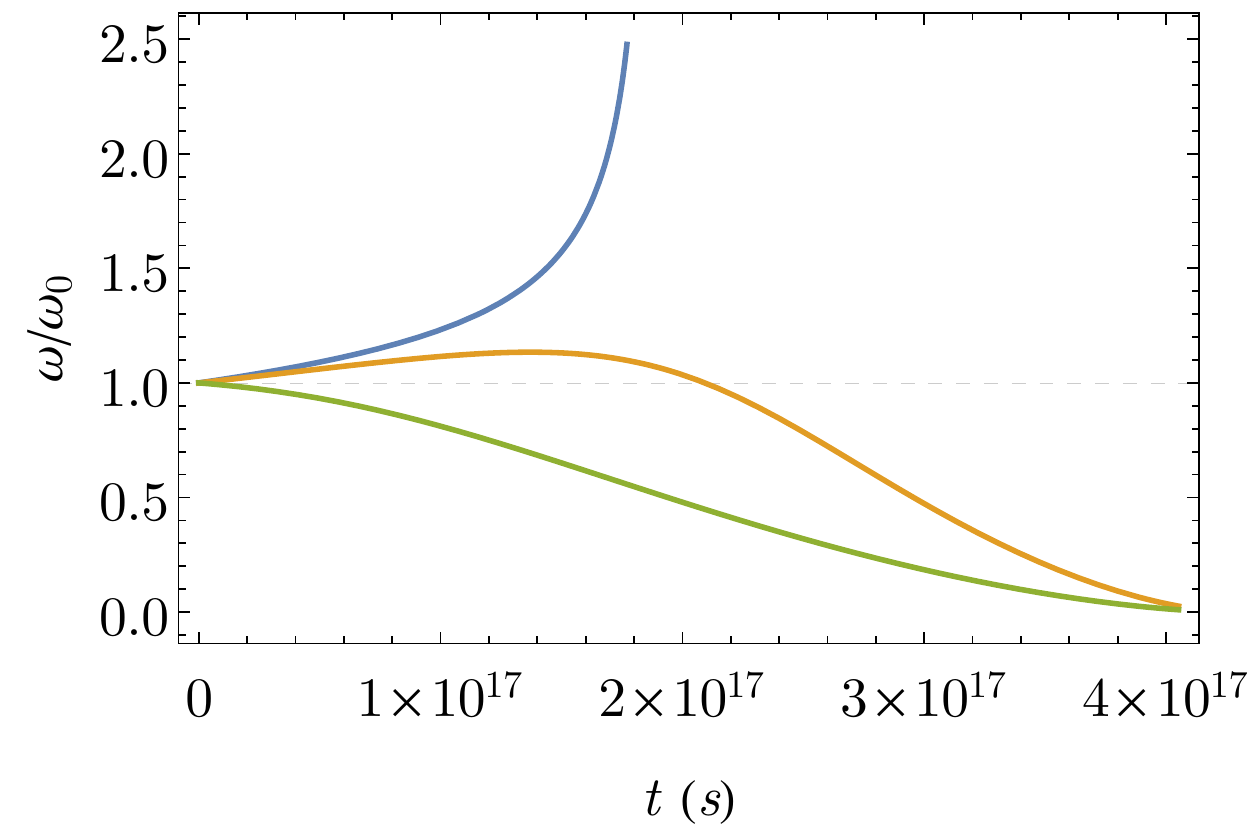}
\caption{\label{fig:freq_equal_mass} Evolution of the orbital frequency $\omega$ (normalized to the initial  orbital frequency $\omega_{0}$) between two evaporating black holes with initial mass $m_{0} = \SI{e12}{kg}$, as a function of time (in seconds). The color of the three different curves correspond to the ones used in Fig. \ref{fig:R_equal_mass}.}
\end{figure}

Since the system is assumed to be Keplerian at every step in its evolution, using Kepler's third law $\omega^{2} = 2Gm/R^{3}$ along with Eq. \eqref{eq:R_analytic} provides an analytical expression for the orbital frequency $\omega$:
\begin{equation}
    \omega(t) = \omega_{0} \left(1 - \frac{t}{t_{\mathrm{ev}}}\right)^{\frac{5}{3}} \left(1 + \frac{1}{6} \frac{t_{\mathrm{ev}}}{t_{\mathrm{cc}}} \left[\left(1 - \frac{t}{t_{\mathrm{ev}}}\right)^{6} -1 \right] \right)^{-\frac{3}{8}}
\label{eq:freq_analytic}
\end{equation}
with $\omega_{0}^{2} \equiv 2Gm_{0}/R_{0}^{3}$ the initial orbital frequency. As for the orbital separation, it admits three distinct regimes (see Fig. \ref{fig:freq_equal_mass}). The frequency of gravitational waves is simply twice the orbital frequency.
 
As expected, when $R \to \infty$ in the outspiralling case, $\omega \to 0$. This basically means that the two bodies are not sufficiently tightly bound to be considered anymore as a binary system, hence the very notion of orbital frequency becomes irrelevant.

When the system actually merges (blue curve in Fig. \ref{fig:freq_equal_mass}), we recover the familiar chirping trend. To determine the chirping frequency, we place ourselves at the innermost stable circular orbit (ISCO) which corresponds to an orbital separation $R_{\mathrm{ISCO}} = 12Gm_{\mathrm{ISCO}}/c^{2}$ where $m_{\mathrm{ISCO}}$ is the time-varying mass evaluated at $t = t_{\mathrm{ISCO}}$. Thus one can solve the equation $R(t) = R_{\mathrm{ISCO}}$ using Eq. \eqref{eq:R_analytic}, find the corresponding time $t_{\mathrm{ISCO}}$ and then evaluate $\omega(t_{\mathrm{ISCO}})$ with Eq. \eqref{eq:freq_analytic}. Since $R(t) = R_{\mathrm{ISCO}}$ is an implicit equation that doesn't admit any simple analytical solution, this should be done numerically. 

Nonetheless, two important features can be noticed. First, as for the standard case, $t_{\mathrm{ISCO}}$ is very close to the effective time of coalescence $t_{\mathrm{coal}}$. Solving $R(t) = R_{\mathrm{ISCO}}$ for various values of $m_{0}$ (typically from $\SI{e9}{kg}$ to $\SI{e20}{kg}$) shows that, if $\gamma$ is the order of magnitude of the initial mass, {\it i.e.} if $m_{0} \simeq 10^{\gamma}$ kg, then $t_{\mathrm{coal}} - t_{\mathrm{ISCO}} \simeq 10^{\gamma-34}$ s. For PBHS, one can thus legitimately approximate $t_{\mathrm{ISCO}} \sim t_{\mathrm{coal}}$. Conveniently, we readily have an explicit expression for $t_{\mathrm{coal}}$, see Eq. \eqref{eq:tau_effective_coal}. However, by construction, $\omega(t_{\mathrm{coal}}) = \infty$. Although tiny in itself, the difference between $t_{\mathrm{ISCO}}$ and $t_{\mathrm{coal}}$ is, in principle, relevant to determine the final frequency. 

Analytically, the expression \eqref{eq:freq_analytic} is not of great use and one should simply go back to Kepler's third law. In fact, whenever the binary system actually merges, the relative variation of mass during the process is small ; more specifically, we shall demonstrate that it is perturbatively small at ISCO. Let us introduce 
\begin{equation}
    \varepsilon \equiv \frac{m_{0} - m_{\mathrm{ISCO}}}{m_{0}} = 1 - \left(1 - \frac{t_{\mathrm{ISCO}}}{t_{\mathrm{ev}}}\right)^{1/3}, \quad \varepsilon >0,
\label{eq:epsilon_parameter}
\end{equation}
where the last equality comes from Eq. \eqref{eq:hawking_mass}. Using that $t_{\mathrm{ISCO}} \sim t_{\mathrm{coal}}$ and Eq. \eqref{eq:tau_effective_coal}, we are led to
\begin{equation}
     \frac{t_{\mathrm{ISCO}}}{t_{\mathrm{ev}}} \sim \frac{t_{\mathrm{coal}}}{t_{\mathrm{ev}}} = 1 - \left[1 - 6\frac{t_{\mathrm{cc}}}{t_{\mathrm{ev}}} \right]^{1/6}.
\label{eq:tisco_and_tcoal}
\end{equation}
Looking at Fig. \ref{fig:t_coal}, unless we put ourselves right at the saturated condition $t_{\mathrm{ev}} = 6 t_{\mathrm{cc}}$ -- this case is treated at the end of the discussion -- we have $t_{\mathrm{ISCO}}/t_{\mathrm{ev}} \lesssim 0.4$ which leads to $\varepsilon \lesssim 0.2 \ll 1$, {\it i.e.} the variation of mass is always small whenever the binary system merges. Consequently, one can rely on the usual expression for the frequency at ISCO, but evaluated for bodies that have lost a fraction $\varepsilon$ of their initial masses such that $m(t_{\mathrm{ISCO}}) = m_{0} (1-\varepsilon)$. If $f_{\mathrm{ISCO}}^{\mathrm{H}}$ (resp. $f_{\mathrm{ISCO}}^{\mathrm{cc}}$) denotes the frequency at ISCO for the evaporating binary system (resp. for a binary system formed of two \textit{constant} masses $m_{0}$), then
\begin{equation}
    \frac{f_{\mathrm{ISCO}}^{\mathrm{H}}}{f_{\mathrm{ISCO}}^{\mathrm{cc}}} = \frac{1}{1-\varepsilon} \sim 1+\varepsilon >1,
\label{eq:ratio_fisco}
\end{equation}
hence for an evaporating binary system, we expect (slightly) higher frequencies. For numerical estimates, plugging Eq. \eqref{eq:epsilon_parameter}, supplemented by Eq. \eqref{eq:tisco_and_tcoal}, into Eq. \eqref{eq:ratio_fisco}, one is led to:
%and using textbooks expressions for $f_{\mathrm{ISCO}}^{\mathrm{cc}}$ \cite{Maggiore:2007ulw, Antelis:2018sfj} gives the approximate formula

%\begin{equation}
%    f_{\mathrm{ISCO}}^{\mathrm{H}} \simeq \SI{2200}{Hz} \left(1- \frac{t_{\mathrm{coal}}}{t_{\mathrm{ev}}}\right)^{-1/3} \frac{M_\odot}{m}
%\end{equation}
%where $M_{\odot}$ is the stellar mass. 

\begin{equation}
\begin{split}
    f_{\mathrm{ISCO}}^{\mathrm{H}} &\simeq\frac{1}{12\pi\sqrt{6}}\left(1- \frac{t_{\mathrm{coal}}}{t_{\mathrm{ev}}}\right)^{-1/3}\frac{c^3}{Gm} \\
    &\simeq \SI{2200}{Hz} \left(1- \frac{t_{\mathrm{coal}}}{t_{\mathrm{ev}}}\right)^{-1/3} \frac{M_\odot}{m},
\end{split}
\end{equation}
where $M_{\odot}$ is the stellar mass. As for the strain amplitude at the ISCO, it is basically given by the usual formula \cite{Antelis:2018sfj}:
\begin{equation}
    h_{\mathrm{max}} \approx \frac{2}{D}\left(\frac{G M_{\mathrm{c}}}{c^{2}}\right)^{5/3} \left(\frac{\pi f_{\mathrm{ISCO}}}{c}\right)^{2/3},
\label{eq:strain_formula}
\end{equation}
where $M_{\mathrm{c}}$ is the chirp mass -- simply given by $M_{\mathrm{c}} = m/2^{1/5}$ in the case of bodies of identical masses -- and $D$ is the distance to the observer. Applying the same reasoning than for the frequency associated with the ISCO, one easily obtains that $h_{\mathrm{max}}^{\mathrm{H}}/ h_{\mathrm{max}}^{\mathrm{cc}} = 1- \varepsilon <1$, {\it i.e.} using again Eqs. \eqref{eq:epsilon_parameter} and \eqref{eq:tisco_and_tcoal}:
\begin{equation}
    h_{\mathrm{max}}^{\mathrm{H}} \simeq \left(1- \frac{t_{\mathrm{coal}}}{t_{\mathrm{ev}}}\right)^{1/3} h_{\mathrm{max}}^{\mathrm{cc}}.
\end{equation}

Let us now examine the specific case for which $t_{\mathrm{ev}} = 6 t_{\mathrm{cc}}$. According to Fig. \eqref{fig:t_coal} -- or equivalently Eq. \eqref{eq:tau_effective_coal} -- it means that the system merges concomitantly to the full evaporation of the two BHs, {\it i.e.} $t_{\mathrm{coal}} = t_{\mathrm{ev}}$. Furthermore, plugging the condition $t_{\mathrm{ev}} = 6 t_{\mathrm{cc}}$ into Eq. \eqref{eq:R_analytic} provides 
\begin{equation}
    R(t) = R_{0} \sqrt{1- \frac{t}{t_{\mathrm{ev}}}},
\end{equation}
which proves that in that case, the system does actually merge. On the other hand, using Eq. \eqref{eq:freq_analytic}, it is easy to see that the frequency is simply given by 
\begin{equation}
    \omega(t) = \omega_{0} \left(1 - \frac{t}{t_{\mathrm{ev}}}\right)^{-7/12}.
\label{eq:freq_saturated}
\end{equation}
We recover that at coalescence, the frequency formally diverges. The analysis of the frequency reached at the ISCO can however be here conducted in a fully analytical way. Solving $R(t) = R_{\mathrm{ISCO}}$ indeed leads to 
\begin{equation}
    t_{\mathrm{ISCO}} = t_{\mathrm{ev}} \left[ 1- \left(\frac{12Gm_{0}}{R_{0}c^{2}}\right)^{6}\right].
\end{equation}
One can plug the above expression into Eq. \eqref{eq:freq_saturated} and use the explicit form of $\omega_{0}$, as well as the condition $t_{\mathrm{ev}} = 6 t_{\mathrm{cc}}$, which enables to express the initial orbital separation $R_{0}$ as a function of the initial mass $m_{0}$ and some constants (or vice versa), to obtain that the frequency associated at the ISCO is given by:
\begin{equation}
    \omega (t_\mathrm{ISCO}) = \left(\frac{c^{3}}{G}\right)^{3/2} \frac{1}{\sqrt{34992 \alpha_{\mathrm{H}}}}.
\label{eq:PBH_ISCO_max_freq}
\end{equation}

Nicely, this formula does depend neither on $m_0$, nor on $R_0$. As the ISCO is reached very late in the process (one should keep in mind that, in this case, the mass decreases during the inspiral and vanishes at merging), the frequency is huge, not far from the Planck frequency. This is obviously a purely academic situations but it enlightens the behaviour of the system in the most extreme (merging) case. Interestingly, one should notice that the binary system always reaches its ISCO, which was not {\it a priori} obvious, as $R_{\mathrm{ISCO}}=12Gm/c^2$ decreases as $R$ decreases. The two curves ($R(t)$ and $R_{\mathrm{ISCO}}(t)$) however necessarily intersect each other.

\section{\label{sec:sec4} Bondi accretion of phantom dark energy}
We now turn to the study of binaries formed of two black holes accreting phantom dark energy. As in Ref. \cite{Mersini-Houghton:2008cul, Enander_2010}, to describe the accretion of matter from interstellar medium by a compact object, we shall use the standard Bondi approximation corresponding to a rate of mass change $\dot{m} \propto - m^{2}$, which integrates into
\begin{equation}
    m(t) = \frac{m_{0}}{1+t/\tau},
\label{eq:Bondi_mass}
\end{equation}
the typical time of evolution $\tau$ reading, in the context of dark energy \cite{Babichev:2004yx, Enander_2010},
\begin{equation}
    \tau = \frac{\SI{3e40}{s}}{\abs{1+w}} \left( \frac{\rho_{\mathrm{d}}}{\rho_{c}}\right)^{-1} \frac{M_{\odot}}{m_{0}},
\end{equation}
where $M_{\odot}$ is the Solar mass, $w=p/\rho$ is the equation of state parameter, $\rho_{\mathrm{c}} \sim \SI{e-26}{kg/m^{3}}$ is the critical density of the universe, and  $\rho_{\mathrm{d}} (\infty)$ is the dark energy density (which is of the same order of magnitude as shown by observations \cite{Planck:2018vyg}). For numerical estimates, we assume in the following $1+w\sim -0.1$.

Several remarks are in order at this point. First, let us emphasize that we focus here on phantom energy ($w<-1$), that is on ``anti-accretion" (accretion inducing a mass decrease) as the case $w>-1$ leads to a standard accretion during which the mass increases. Although interesting in itself, this situation does not bring any significant new features to the evolution of the orbital separation: both the effects of gravitational waves and of accretion play in the same direction. The case of a pure cosmological constant, that is $w=-1$, leads to no mass variation at all.

Second, it should be emphasized that in the case of a standard accretion, as it will be discussed later in this article, Eq. \eqref{eq:Bondi_mass} would lead to a mass divergence in a {\it finite} amount of time. Although not directly related to this work, this opens interesting phenomenological features \cite{Barrau:2022bfg}. The situation is, in a sense, formally close -- although reversed -- to the one of the Hawking evaporation. The deep reason for this is quite simple. In the case of Hawking evaporation, as in the case of standard accretion, the mass variation gets amplified by the evolution it generates. When it evaporates, a BH becomes smaller and smaller, hence hotter and hotter. The phenomenon diverges in finite time. Exactly as what happens for standard accretion: the more a BH absorbs usual matter (with $w>-1$), the larger the cross section gets and the faster its mass grows, leading to a divergence in finite time. Both cases are fundamentally unstable and the relevant question is basically to wonder if the coalescence (if any) happens before or after the singularity associated with the mass variation. On the other hand, the case of phantom energy accretion is stable, hence the regular behaviour of Eq. \eqref{eq:Bondi_mass}, whatever the value of $t$. This is because, when $w<-1$, as the anti-accretion takes place, the BH gets smaller and smaller and, therefore, sees its cross section and, consequently its mass loss rate, {\it decrease} with time. It is a negative feedback whereas the diverging cases correspond to positive feedbacks.

\subsection{Evolution of the orbital separation}
Focusing on the case of phantom dark energy, the same method as used in section \ref{sec:sec3} can be applied. The differential equation now reads
\begin{equation}
    \dot{R} = - \frac{128}{5}\frac{G^{3}}{c^{5}} \frac{m^{3}}{R^{3}} + \frac{3}{\tau}\frac{m}{m_0} R, 
\label{eq:diff_bondi}
\end{equation}
which clearly shows that, depending on an initial setup favouring either inspiral or outspiral, there is always one of the two terms which presents a competitive behaviour between the time evolution of the mass and its orbital separation counterpart (in addition to the obvious competition between the two terms themselves, simply due to the opposite sign). Explicitly: if $R$ decreases, the evolution of the amplitude of the first term is not obvious as $m$ also decreases, making the fate of $m^3/R^3$ {\it a priori} non-trivial. If, the other way round, $R$ increases, the evolution of the amplitude of the second term is now not obvious as $m$ still decreases, making the evolution of $mR$ possibly intricate.

\begin{figure}[h]
\centering
\includegraphics[width=0.48 \textwidth]{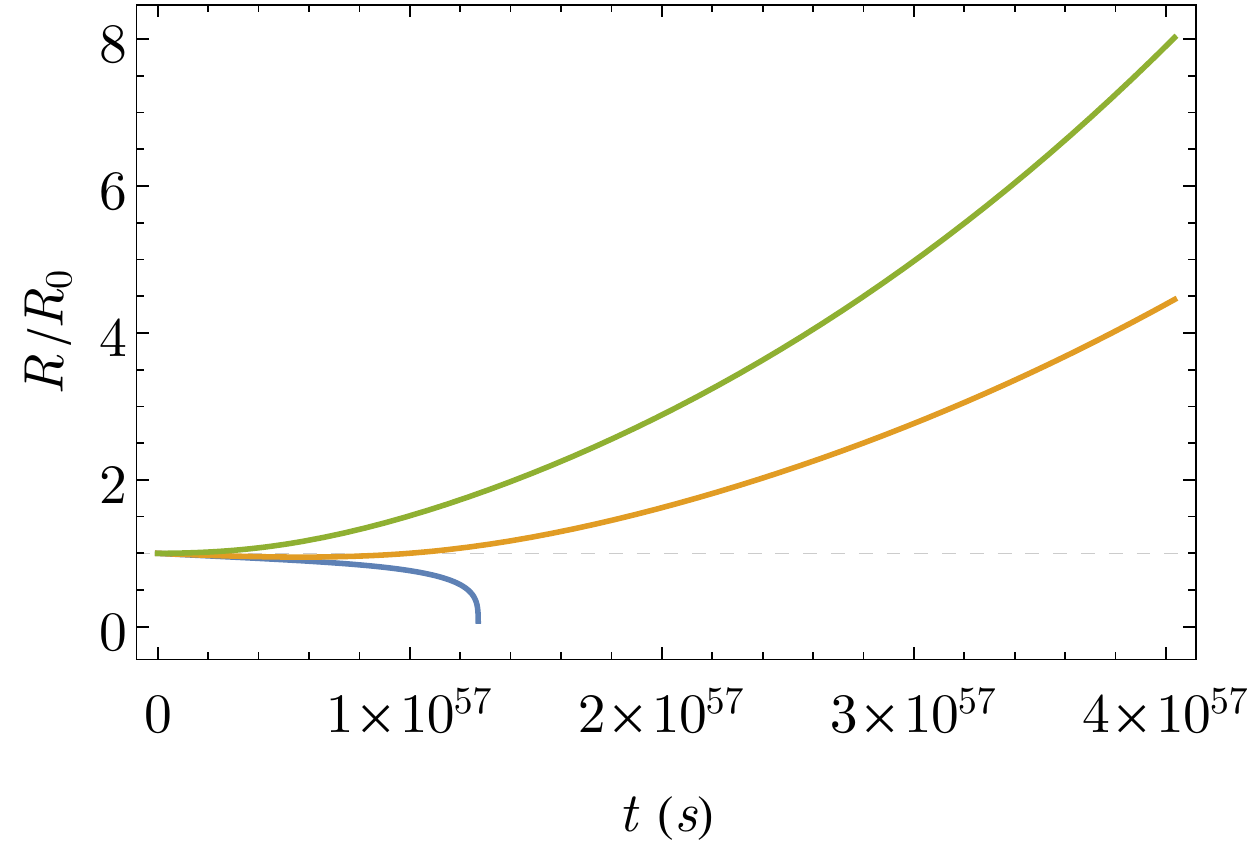}
\caption{\label{fig:R_Bondi} Evolution of the orbital separation $R$ (normalized to the initial orbital separation $R_{0}$) between two black holes submitted to a Bondi-type accretion of phantom dark energy with initial mass $m_{0} = \SI{e12}{kg}$, as a function of time (in seconds). We recover the outspiralling dynamics for $\tau < 12t_{\mathrm{cc}}$ (green curve), the inspiralling one for $\tau > 14 t_{\mathrm{cc}} $ (blue curve), and the intermediate non-monotonic regime for $12 t_{\mathrm{cc}} < \tau < 14t_{\mathrm{cc}} $ (orange curve).}
\end{figure}

At an initial time $t_{0} = 0$, one has
\begin{equation}
    \frac{\dot{R}}{R} \bigg\rvert_{t_{0} = 0} = - \frac{1}{4t_{\mathrm{cc}}} + \frac{3}{\tau},
\end{equation}
and the integration of Eq. \eqref{eq:diff_bondi} provides 
\begin{equation}
      R(t) = R_{0} \left(1+ \frac{t}{\tau} \right)^{3} \left(1 +\frac{1}{14} \frac{\tau}{t_{\mathrm{cc}}} \left[ \left(1 + \frac{t}{\tau} \right)^{-14} - 1 \right] \right)^{1/4}.
\end{equation}
Solving the equation $R(t) = 0$ gives the following effective time of coalescence
\begin{equation}
    t_{\mathrm{coal}} = \tau \left( \left[1 - 14 \frac{t_{\mathrm{cc}}}{\tau}\right]^{-1/14} -1\right),
\label{eq:Bondi_tcoal}
\end{equation}
which is well-defined mathematically only for $\tau > 14 t_{\mathrm{cc}}$. As a consequence, the system outspirals for $\tau < 12 t_{\mathrm{cc}}$, inspirals when $\tau > 14 t_{\mathrm{cc}}$ and for $12 t_{\mathrm{cc}} < \tau < 14 t_{\mathrm{cc}}$ we recover this so-called intermediate regime beginning with an initial inspiral but with an ultimate outspiral due to the time-varying mass. Contrary to the case of Hawking evaporation though, the outspiralling dynamics does not lead to any divergence of the orbital separation at finite time, as it can be seen in Fig. \ref{fig:R_Bondi}. As the evolution law for the mass -- Eq. \eqref{eq:Bondi_mass} -- does not present any pole, the orbital radius gently tends to infinity for $t\to \infty$.

\subsection{Evolution of the frequency}
\begin{figure}[h]
\centering
\includegraphics[width=0.48 \textwidth]{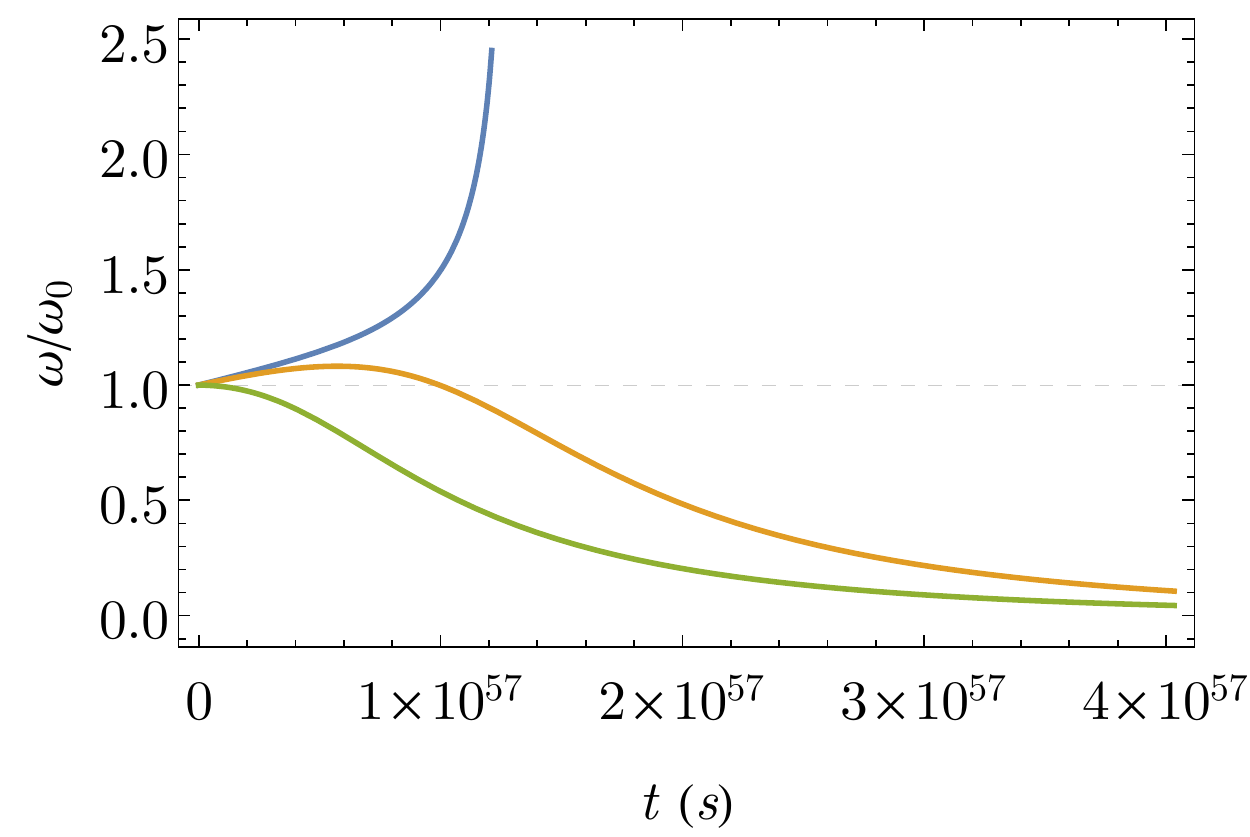}
\caption{\label{fig:freq_Bondi} Evolution of the orbital frequency $\omega$ (normalized by the initial  orbital frequency $\omega_{0}$) between two black holes submitted to Bondi-type accretion of phantom dark energy with initial mass $m_{0} = \SI{e12}{kg}$, as a function of time (in seconds). The color of the three different curves correspond to the ones used in Fig. \ref{fig:R_Bondi}.}
\end{figure}

It is straightforward, using Kepler's third law, to obtain an analytical expression for the orbital frequency. Its evolution is shown in Fig. \ref{fig:freq_Bondi}. The ISCO analysis performed in the case of the Hawking evaporation is still applicable. One can indeed demonstrate that $t_{\mathrm{ISCO}} \sim t_{\mathrm{coal}}$ and plotting the curve $t_{\mathrm{coal}}/\tau$ using Eq. \eqref{fig:R_Bondi} shows that the previously introduced $\varepsilon$ parameter is such that $\varepsilon \lesssim 0.23$. Then an approximate formula for the frequency associated to the ISCO when the system is in the inspiralling regime is
%\begin{equation}
%    f^{\mathrm{B}}_{\mathrm{ISCO}} \simeq \SI{2200}{Hz} \left(1 +  \frac{t_{\mathrm{coal}}}{\tau}\right) \frac{M_\odot}{m}
%\end{equation}

\begin{equation}
\begin{split}
    f_{\mathrm{ISCO}}^{\mathrm{B}} &\simeq\frac{1}{12\pi\sqrt{6}}\left(1+ \frac{t_{\mathrm{coal}}}{\tau}\right)\frac{c^3}{Gm} \\
    &\simeq\SI{2200}{Hz} \left(1 +  \frac{t_{\mathrm{coal}}}{\tau}\right) \frac{M_\odot}{m},
\end{split}
\end{equation}
with the time of coalescence $t_{\mathrm{coal}}$ being given by Eq. \eqref{eq:Bondi_tcoal}. Again, the correction brought by the time-varying mass is meager. Carrying out the same analysis, but with a \textit{non}-phantom dark energy, {\it i.e.} with $w>-1$, shows, as expected, that the binary system always merges. In addition, because the mass growth enhances the inspiralling effect, it does so with a shorter time of coalescence. However, as we shall show later on, for all ranges of mass, the difference between these two times of coalescence is too small in practice to constitute a reliable experimental criteria to differentiate the case $w>-1$ from the case $w<-1$.

\subsection{Strain - observational distance relation}
We now consider the possibility, in the light of our results and in the same vain as in Ref. \cite{Enander_2010}, to use observations of a binary system to measure local dark energy properties. To get an order of magnitude one can safely assume that, for the accretion of dark energy not to be negligible, the two terms entering the differential equation \eqref{eq:diff_bondi} must be of the same order of magnitude. This implies that the two bodies should be then separated by (at least) a distance 
\begin{equation}
    R \sim \left(\frac{128}{15} \frac{G^{3}}{c^{5}} \tau m_{0}\right)^{1/4} \sqrt{m}.
\label{eq:equal_distance}
\end{equation}
Using Kepler's third law and the above relation to express the frequency solely in function of the mass, and injecting the result in the strain formula, Eq. \eqref{eq:strain_formula}, shows that, given a minimum detectable strain amplitude $h_{\mathrm{min}}$, and demanding $h> h_{\mathrm{min}}$, constrains the distance to the observer $D$ to be less than 
\begin{equation}
    D_{\mathrm{max}} \approx \SI{2.9e-2}{pc} \left(\frac{m}{M_{\odot}}\right)^{3/2} \left(\frac{h_{\mathrm{min}}}{10^{-23}}\right)^{-1},
\label{eq:max_distance}
\end{equation}
where the normalizing value $10^{-23}$ corresponds to the order of magnitude of the minimum measurable strain amplitude currently reached by gravitational wave detectors (see {\it e.g.} \cite{Virgo:2022ysc}). This expression depends only on the mass as the orbital separation is chosen so that the anti-accretion term plays a significant role. It should be stressed that it is very different -- not only in prefactors but also in the mass functional dependence -- from the usual formula encountered in gravitational wave physics, $D<1.6(m/M_{\odot})/(h\times 10^{23})$ Gpc, as the latter assumes that the system is seen at the merging time. This is far from being the case for the situation we consider here.

Beyond the distance $D_{\mathrm{max}}$, even if the radial change due to the accretion of dark energy were big when compared to the one due to the emission of GWs, the strain produced would be too small  to be detected. For solar-mass BHs, with $h_{\min} \sim 10^{-23}$, the corresponding order of magnitude is approximately the size of the solar system. However, for supermassive black holes with $m=10^{8} M_{\odot}$, Eq. \eqref{eq:max_distance} gives  $D_{\mathrm{max}} \sim \SI{29}{Gpc}$ which is more than the cosmological horizon radius corresponding to the limit of the observable universe. It therefore seems that wherever a binary system of $10^{8} M_{\odot}$ BHs, seen early enough in the inspiralling process, exists, the accretion of phantom dark energy could be measured on Earth. This is however not that simple. 

Whenever phantom dark energy dominates over gravitational radiation, the binary system is in an outspiralling regime, the orbital frequency thus decreases (see the orange and green curves in Fig. \ref{fig:freq_Bondi}). 
%For solar-mass size BHs or above -- thus ruling out any lighter BHs submitted to Hawking radiation -- this could be a relevant experimental observation to ascertain that the binaries under scrutiny are driven by accretion of phantom dark energy : the decreasing evolution of the frequency would indeed definitely differs from the typical chirping trend. Furthermore, for heavy BHs, the range of frequencies could be not too small to be detected. However, one is left with two unavoidable drawbacks : first Eq. \eqref{eq:max_distance} constrains the mass of these hypothetical BHs to supermassive ones so that the exploitable strain amplitude is compatible with the maximum distance of observation. 
This is an appealing feature as it never happens in standard cases corresponding to a dynamics driven only by the emission of GWs. It could be a ``smoking gun" for an effect beyond the standard model (BSM). For the strain to be high enough to be detected without requiring the system to violate obvious bounds (as it would, {\it e.g.}, be the case if one assumes the existence solar mass BHs within the solar system), the best candidates are supermassive BHs.
Equation \eqref{eq:equal_distance} shows that they should be separated by a distance of order $R \sim \SI{e18}{m}$, corresponding to a time of merging above the Hubble time. However, although the strain would actually be measurable and the sign of the frequency drift would be favourable (that is negative and, therefore, distinct from any usual phenomenon), the amplitude of the  \textit{variation} of frequency -- which stands as the relevant parameter -- would be ridiculously small. Fixing orders of magnitude by assuming  constant masses, one would have $\lvert \dot{f}/f\rvert \sim 10^7 f^{8/3}$ which leads to $\lvert \dot{f}/f \rvert \sim 10^{-30}$ for $m = 10^{8} M_{\odot}$. By inspection of Fig. \ref{fig:freq_Bondi}, it is obvious that taking into account the variation of the mass and the precise form of the frequency would only worsen the effect. In addition, the absolute value of the frequency would also be way below anything measurable in the next decades.

It should be noticed that, in the context of binaries of evaporating BHs, the question of the maximum distance of observation was already considered in Ref. \cite{Barrau:2023kuv}: BHs undergoing the Hawking effect (with masses typically below $M_{*} \equiv \SI{e12}{kg}$) would have fully evaporated in a time smaller than the age of the Universe. There is no chance to observe them in our necessarily confined scope of observation. 
%Besides, this paper shows that the Hawking radiation would not significantly enhance the frequency associated with the ISCO, nor the strain.
In this case, the phenomenological relevance of our work is not about the detection of a single event but, rather, about the statistical features of the  expected merging rate.

\section{\label{sec:sec5} Generic results}
Beyond the examples of (potential) phenomenological interest given in the previous sections, we now focus on offering an exhaustive classification of the possible behaviours of binaries emitting GWs submitted to mass variation effects described by either a positive or a negative power law. The analytical solutions we give -- not previously known in the relevant literature to the best of our knowledge -- can be used in many different situations. We consider mass loss rates of the form:
\begin{equation}
\begin{split}
    &(i) \quad \dot{m} = - \frac{\beta}{m^{k-1}}, k> 0,\\
    &(ii) \quad \dot{m} = - \beta m^{k+1}, k >0,\\   
    &(iii) \quad \dot{m} = \beta m^{k+1}, k >0,\\
    &(iv) \quad \dot{m} = \frac{\beta}{m^{k-1}}, k>0,
\end{split}
\end{equation}
with $\beta > 0$ in all cases. For now, we ignore the cases for which the above differential equations do not lead to a mass evolution following a power law -- that is corresponding to $\dot{m} \propto m$ which implies an exponential trend. Those specific types of evolution are dealt with at the end of this section.

Let us comment about the physical signification of some particular cases. Hawking radiation corresponds to $(i)$ with $k=3$. A constant mass loss rate -- as it would be the case for a star during most of its life -- corresponds to $(i)$ with $k=1$. A standard Bondi accretion by a black hole -- that is to say a spherically symmetric, adiabatic, steady-flow gas accretion -- corresponds to $(iii)$ with $k=1$. A Bondi anti-accretion -- that is with a phatom fluid -- corresponds to $(ii)$ with $k=1$. A naive mass loss rate proportional to the area of the compact (non-black hole) object corresponds to $(ii)$ with $k=1/3$. A hypothetical situation where the accretion rate of a BH would be solely driven by its surface gravity corresponds to $(iv)$ with $k=2$.

%The case $(i)$ corresponds to a mass loss through an evaporation process, such as Hawking radiation (which corresponds to $k=2$). The cases $(ii)$ and $(iii)$ correspond to both accretion phenomena. But in $(iii)$, it is pure accretion, that is to say, the mass increases with time thanks of the absorption of matter for instance. $(ii)$ is however a situation where the mass decreases but by radiating for instance. For $k=1$ in $(ii)$ and $(iii)$, it represents Bondi accretion process, that is to say a spherically symmetric, adiabatic, steady-flow gas accretion, first applied to stars and in $(ii)$ it is an accretion of a phantom fluid with $w <-1$. The case $(iv)$ has less physical relevance, but we shall also solve it for sake of completeness. 

Each of these differential equations for $\beta$ constant are separable and easily integrate into :
\begin{equation}
\begin{split}
    &(i) \quad m(t) = m_{0} \left(1 - \frac{t}{t_{\mathrm{ev}}}\right)^{1/k}, \quad t_{\mathrm{ev}} = \frac{m_{0}^{k}}{k\beta} >0, \\
    &(ii) \quad m(t) = m_{0} \left(1 + \frac{t}{\tau}\right)^{-1/k}, \quad \tau = \frac{1}{m_{0}^{k}k\beta} > 0, \\   
    &(iii) \quad m(t) = m_{0} \left(1 - \frac{t}{\tau_{*}}\right)^{-1/k}, \quad \tau_{*} = \frac{1}{m_{0}^{k}k\beta} > 0, \\
    &(iv) \quad m(t) = m_{0} \left(1 + \frac{t}{\tilde{\tau}}\right)^{1/k}, \quad \tilde{\tau} = \frac{m_{0}^{k}}{k\beta} >0.
\end{split}
\label{eq:generic_mass}
\end{equation}
It should be noticed that $(ii)$ and $(iv)$ exhibit behaviours qualitatively different from $(i)$ and $(iii)$. In the latter two cases, there is a critical point in the evolution of the mass: for $(i)$ it corresponds to $t=t_{\mathrm{ev}}$, where the mass suddenly plunges to zero whereas for $(iii)$ it corresponds to the divergence at finite time $t=\tau_{*}$. As previously explained, those cases are physically related to unstable processes driven by positive feedbacks. Conversely, the cases $(ii)$ and $(iv)$ correspond, respectively, to a smooth decrease and a smooth increase of the mass. They are related to stable processes with negative feedbacks. Then, $\tau$ and $\tilde{\tau}$ can be understood as typical time scales on which the considered body has lost or gained a significant amount of mass. 

The differential equations satisfied by the orbital separation $R$ are respectively given by :
\begin{equation}
\begin{split}
    &(i) \quad \dot{R} = - \frac{128}{5} \frac{G^{3}}{c^{5}} \frac{m_{0}^{3}}{R^{3}}\left(1 - \frac{t}{t_{\mathrm{ev}}}\right)^{3/k} + \frac{3}{k} \frac{R}{t_{\mathrm{ev}} - t}, \\
    &(ii) \quad \dot{R} = - \frac{128}{5} \frac{G^{3}}{c^{5}} \frac{m_{0}^{3}}{R^{3}}\left(1 + \frac{t}{\tau}\right)^{-3/k} + \frac{3}{k} \frac{R}{\tau + t}, \\
    &(iii) \quad \dot{R} = - \frac{128}{5} \frac{G^{3}}{c^{5}} \frac{m_{0}^{3}}{R^{3}}\left(1 - \frac{t}{\tau_{*}} \right)^{-3/k} - \frac{3}{k} \frac{R}{\tau_{*} - t}, \\
    &(iv) \quad \dot{R} = - \frac{128}{5} \frac{G^{3}}{c^{5}} \frac{m_{0}^{3}}{R^{3}}\left(1 + \frac{t}{\tilde{\tau}}\right)^{3/k} - \frac{3}{k} \frac{R}{\tilde{\tau} + t}, 
\end{split}
\label{eq:Diffs}
\end{equation}
which are all Bernoulli equations. The solutions to these four differential equations are given by :
\begin{align}
    (i) \quad  R(t) = R_{0} &\left( \frac{t_{\mathrm{ev}}}{t_{\mathrm{ev}} -t}\right)^{3/k} \cross \nonumber \\
    &\left(1 + \frac{k}{15+k} \frac{t_{\mathrm{ev}}}{t_{\mathrm{cc}}} \left[ \left(1 - \frac{t}{t_{\mathrm{ev}}} \right)^{\frac{15+k}{k}} - 1 \right] \right)^{1/4}, \nonumber
\\[\jot]
    (ii) \quad R(t) = R_{0}& \left(1+ \frac{t}{\tau} \right)^{3/k} \cross \nonumber \\
    &\left(1 - \frac{k}{k-15} \frac{\tau}{t_{\mathrm{cc}}} \left[ \left(1 + \frac{t}{\tau} \right)^{\frac{k-15}{k}} - 1 \right] \right)^{1/4},\nonumber
\\[\jot]
    (iii) \quad  R(t) = R_{0} &\left(1 - \frac{t}{\tau_{*}} \right)^{3/k} \cross \nonumber \\
    &\left(1 + \frac{k}{k-15} \frac{\tau_{*}}{t_{\mathrm{cc}}} \left[ \left( \frac{\tau_{*}}{\tau_{*} -t}\right)^{\frac{15-k}{k}} - 1 \right] \right)^{1/4},\nonumber
\\[\jot]
    (iv) \quad R(t) = R_{0} &\left(\frac{\tilde{\tau}}{t + \tilde{\tau}} \right)^{3/k} \cross \nonumber \\
    &\left(1 - \frac{k}{15+k} \frac{\tilde{\tau}}{t_{\mathrm{cc}}} \left[ \left( 1 + \frac{t}{\tilde{\tau}} \right)^{\frac{15+k}{k}} - 1 \right] \right)^{1/4}.
\label{eq:separation_generic}
\end{align}
The times of coalescence are given by
\begin{equation}
\begin{split}
    &(i) \quad t_{\mathrm{coal}} = t_{\mathrm{ev}} \left(1 - \left[ 1 - \frac{15+k}{k} \frac{t_{\mathrm{cc}}}{t_{\mathrm{ev}}}  \right]^{\frac{k}{15+k}} \right),\\
    &(ii) \quad t_{\mathrm{coal}} = \tau \left(\left[ 1 - \frac{15-k}{k} \frac{t_{\mathrm{cc}}}{t_{\mathrm{ev}}}  \right]^{\frac{k}{k-15}} -1\right),\\
    &(iii) \quad t_{\mathrm{coal}} = \tau_{*} \left( 1-\left[1 + \frac{15-k}{k} \frac{t_{\mathrm{cc}}}{\tau_{*}} \right]^{\frac{k}{15-k}}\right)\mathrm{for}~k<15,\\
    &~~~~~\quad t_{\mathrm{coal}} = \tau_{*}~\mathrm{for}~k\ge 15,\\
    &(iv) \quad t_{\mathrm{coal}} = \tilde{\tau} \left(\left[1 + \frac{15+k}{k} \frac{t_{\mathrm{cc}}}{\tilde{\tau}} \right]^{\frac{k}{15+k}}-1\right).
\end{split}
\label{eq:coal_generic}
\end{equation}

\begin{figure}[h]
\subfloat[Orbital separation in function of time (in seconds)\label{sfig:testa}]{%
  \includegraphics[width = 0.48 \textwidth]{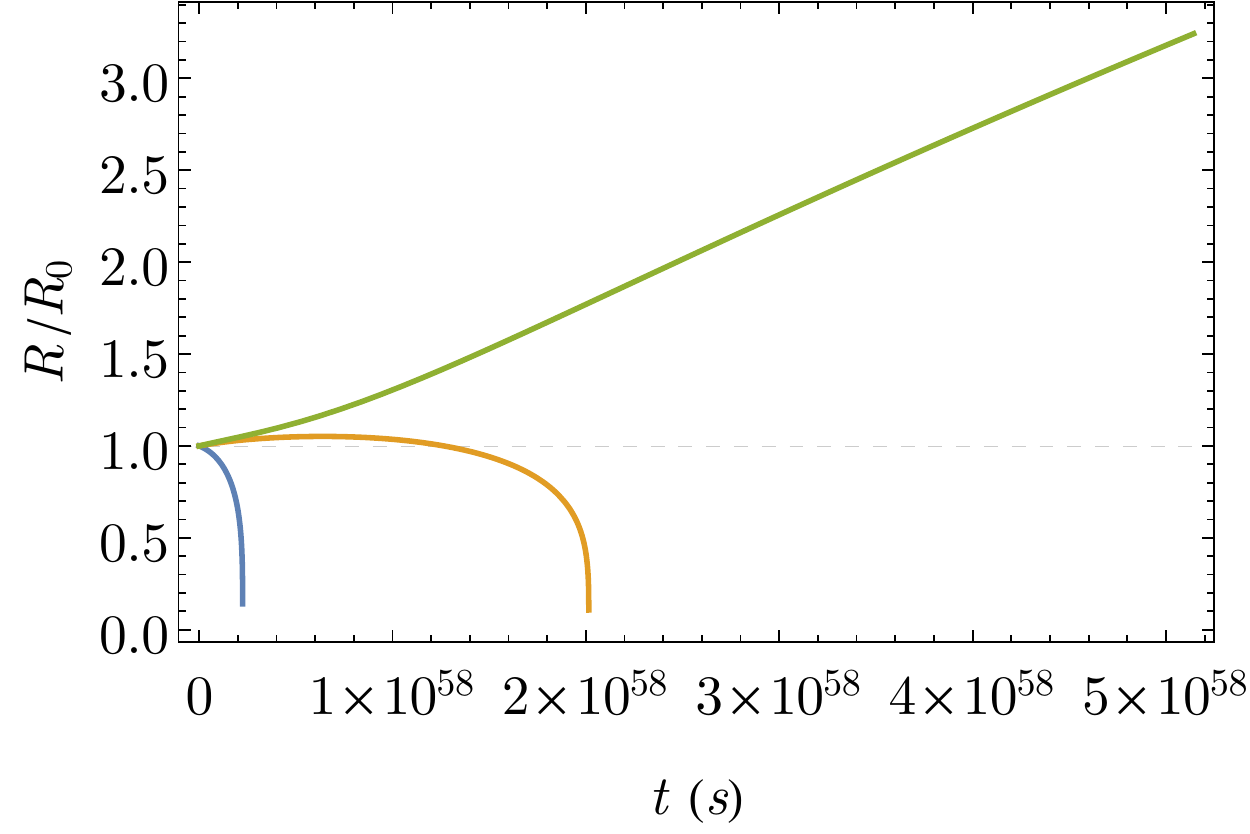}%
}\hfill
\subfloat[Frequency in function of time (in seconds)\label{sfig:testa}]{%
  \includegraphics[width = 0.48 \textwidth]{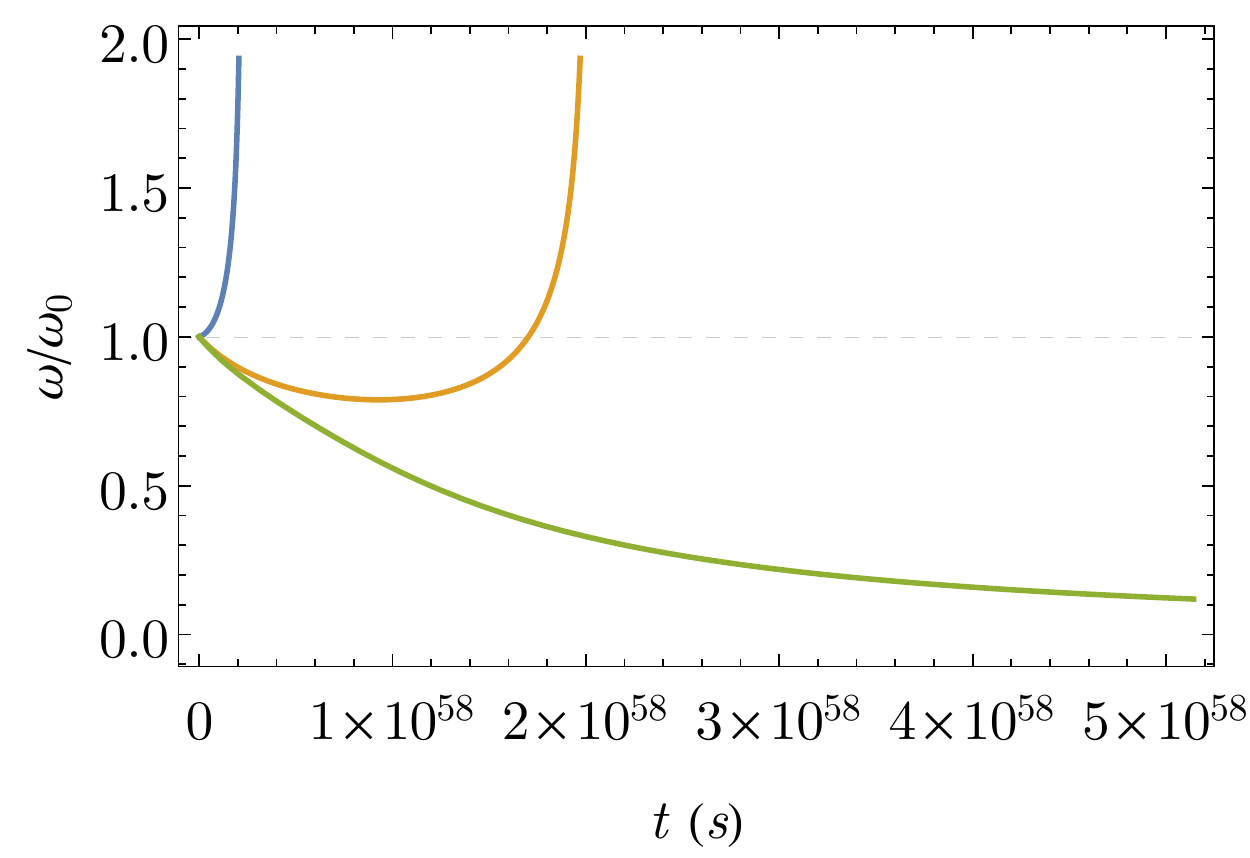}%
}
\caption{Illustration of the behaviour for the case $(ii)$ with $k > 3$. The plot corresponds to $k = 4$, with initial mass $m_{0} = \SI{e12}{kg}$. Panel (a) shows the orbital separation $R$ (normalized to the initial orbital separation $R_{0}$) while panel (b) displays the corresponding orbital frequency. For $\tau < \frac{11}{4} t_{\mathrm{cc}}$ the binary system outspirals (green curve) whereas for $\tau > 3 t_{\mathrm{cc}}$ it inspirals (blue curve). The regime $\frac{11}{4} t_{\mathrm{cc}} <\tau < 3 t_{\mathrm{cc}}$ presents a non-monotonic evolution with an initial outspiral followed by an inspiral when the gravitational radiation takes the lead and makes the binary coalesce (orange curve).}
\label{fig:figs_PDE}
\end{figure}

The different situations are deeply nonequivalent. 
\begin{itemize}
\item In the case $(i)$, the system outspirals whenever $t_{\mathrm{ev}} < \frac{12}{k} t_{\mathrm{cc}}$. It first inspirals and, then, outspirals, for $\frac{12}{k}t_{\mathrm{cc}}<t_{\mathrm{ev}}<\frac{15+k}{k}t_{\mathrm{cc}}$. It only inspirals when $t_{\mathrm{ev}}>\frac{15+k}{k}t_{\mathrm{cc}}$.
\item The case $(ii)$ has to be dived into two sub-cases: 
\begin{itemize}
    \item If $k<3$, the system behaves as previously, but with different bounds. It outspirals for $\tau < \frac{12}{k} t_{\mathrm{cc}}$. It first inspirals and, then, outspirals, for $\frac{12}{k}t_{\mathrm{cc}}<\tau<\frac{15-k}{k}t_{\mathrm{cc}}$. It only inspirals when $\tau>\frac{15-k}{k}t_{\mathrm{cc}}$. 
    \item If, however, $k>3$, the system outspirals for $\tau <\frac{15-k}{k}t_{\mathrm{cc}}$. It first outspirals and, then, inspirals for $\frac{15-k}{k}t_{\mathrm{cc}}<\tau<\frac{12}{k}t_{\mathrm{cc}}$. It inspirals with a monotonic behaviour for $\tau>\frac{12}{k}t_{\mathrm{cc}}$. This new behaviour was not encountered in the phenomenological cases presented before. It is illustrated in Fig. \ref{fig:figs_PDE} with $k=4$ (which would correspond to a mass loss rate proportional to $T^4$, with the temperature $T$ proportional to $m^{5/4}$ as expected in some stellar models).
    \item If $k=3$, there is no intermediate regime. The system outspirals for $4t_{\mathrm{cc}}<\tau$ and inspirals for $4t_{\mathrm{cc}}>\tau$.
\end{itemize}
\item In the cases $(iii)$ and $(iv)$, there is only one regime corresponding to a pure inspiralling behavior leading to the coalescence of the binary system. This was expected since a gain of mass enhances the effect due to gravitational waves. In the case $(iii)$, the mass diverges in a finite amount of time but the system coalesces anyway, either before (when $k<15$) or precisely at this time (when $k\ge 15$). This is not trivial and this results from the fact that the very rapidly growing mass also speeds up the inspiral.
\end{itemize}

The richness of the solutions encountered lies in the asymmetrical role played by the different terms entering the differential equation: the mass variation can either tend to decrease or increase the orbital separation (depending on the sign of $\dot{m}$) and can either be smooth of catastrophic (depending on the exponent of $m$) but the GW term always plays in the same direction, although with a strength coupled to the mass loss.

Let us now turn to the case where $\dot{m} = \pm \beta m, \beta >0$ which leads to a mass evolution exponentially decreasing or increasing, depending on the sign. The above differential equation  integrates into
\begin{equation}
    m_{\pm}(t) = m_{0}e^{\pm t/t_{*}}, \quad t_{*} = \beta^{-1}.
\end{equation}
The differential equation for the orbital separation simply reads
\begin{equation}
    \dot{R} = - \frac{128}{5} \frac{G^{3}}{c^{5}} \frac{m_{0}^{3}}{R^{3}} e^{\pm t/t_{*}} \pm \frac{3}{t_{*}}R
\end{equation}
and integrates into these two solutions (again, depending on the initially chosen sign):
\begin{equation}
    R_{\pm}(t) = R_{0} e^{\pm 3 t/t_{*}} \left(1 \pm \frac{1}{15} \frac{t_{*}}{t_{\mathrm{cc}}} e^{\mp 15 t/t_{*}} \right)^{1/4}.
\label{eq:Expo_R}
\end{equation}
In the case $R_{+}$, The dynamics is always outspiralling. The exponential accretion overwhelms any effect coming from gravitational radiation. It can be clearly seen analytically by taking the limit $t \to \infty$ in Eq. \eqref{eq:Expo_R} for which the term in brackets vanishes and at large $t$, the overall dynamics is dictated by $e^{3t/t_{*}}$. Furthermore, it is straightforward to notice that the equation $R(t) = 0$ does not admit any mathematically sound solution. Conversely, for $R_{-}$, the binary system always merges with a time of coalescence given by
\begin{equation}
    t_{\mathrm{coal}} = \frac{t_{*}}{15} \ln \left(\frac{15 t_{\mathrm{cc}}}{t_{*}} \right).
\end{equation}
As it could have been expected, in both cases, the decreasing or increasing exponential evolution of the mass completely overwhelms the effects coming from gravitational radiation. Although mathematically rigorous, this case of exponential evolution of the mass might suffer from the fact that the formal expression of the radiated power by gravitational waves should be modified (beyond the simple accounting for the $m(t)$ term) when one allows a time-dependent mass, notably including terms related to the rate of variation of the mass. This point is discussed in the following section.

\section{ \label{sec:sec6} Radiated GW power and circularity condition}
In section \ref{sec:sec2}, we have taken into account the radiated power of emitted gravitational waves by substituting in the standard formula a varying mass. One should however be careful about the fact that the standard formula for $P_{\mathrm{gw}}$ is derived with the assumption of a constant mass. For circular orbits and time-varying masses, with orbital frequency $\omega$, the second mass moments $M^{ij} = \mu x^{i}(t) x^{j}(t)$ read:
\begin{equation}
\begin{split}
    &M_{11}(t) = m(t) R^{2} \frac{1-\cos(\omega t)}{4}, \\  &M_{22}(t) = m(t) R^{2} \frac{1+\cos(\omega t)}{4}, \\ &M_{12}(t) = - \frac{1}{4} m(t) R^{2}\sin(\omega t).
\end{split}
\label{eq:secon_mass_moments}
\end{equation}
since for identical masses, the reduced mass is given by $\mu = m/2$. The total radiated power due to the emission of gravitational waves involves the third temporal derivatives of the second mass moments \eqref{eq:secon_mass_moments} since it is given, in the quadrupole approximation, by \cite{Maggiore:2007ulw}
\begin{equation}
    P_{\mathrm{gw}} = \frac{G}{5c^{5}} \langle \dddot{M}_{ij} \dddot{M}_{ij} - \frac{1}{3} (\dddot{M}_{kk})^{2} \rangle,
\label{eq:Pgw_theoretical}
\end{equation}
where the average $\langle \cdot \rangle$ is in the time domain  over several characteristic periods of the gravitational waves. Using Eq. \eqref{eq:secon_mass_moments}, we get
\begin{widetext}
\begin{equation}
P_{\mathrm{gw}} = \frac{G R^{4}}{10c^{5}} \Bigl< \frac{\dddot{m}^{2}}{3} + \frac{3}{2} \left(\frac{3}{2}\ddot{m}^{2} - \dddot{m}\dot{m} \right) \omega^{2} + \frac{3}{2} \left( \frac{3}{2}\dot{m}^{2} -\ddot{m}m \right) \omega^{4} - \dddot{m}m \omega^{3} \sin (\omega t) \cos (\omega t) + \frac{1}{4} m^{2} \omega^{6} \Bigr>.
\label{eq:Pgw_full}
\end{equation}
\end{widetext}
If only the last term in the average in Eq. \eqref{eq:Pgw_full} is taken into account, one recovers the formula \eqref{eq:Pgw}. All the other terms involve derivatives (up to order three) of the mass and are thus the direct consequence of allowing time-varying masses. If $m^{(n)} \equiv \dv[n]{m}{t}$ denotes the $n$-th derivative of the mass, with $n$ a positive integer, then these corrective terms can be neglected if
\begin{equation}
    \forall n, \quad \vert m^{(n)}(t) \vert \ll m \omega^{n},
\label{eq:constraint_general}
\end{equation}
which imposes that the mass of the bodies is slowly varying in comparison to the typical orbital evolution of the binary system. Then, over the temporal window on which the average is performed, that is to say on a few characteristic periods of the gravitational waves, the mass remains nearly constant equal to $m(t)$.

The second assumption used to compute the radiated GW power thanks to Eqs. \eqref{eq:secon_mass_moments} and \eqref{eq:Pgw_theoretical} relies on the fact that the orbit remains circular. In the standard case of bodies with constant mass, the circularity is ensured by demanding $\abs{\dot{\omega}} \ll \omega^{2}$ \cite{Maggiore:2007ulw}. We investigate if this condition is changed by the variability of the mass. By taking the derivative of Kepler's third law, one has
\begin{equation}
     \dot{R} = - \frac{2}{9} (\omega R) \frac{\dot{\omega}}{\omega^{2}} + \frac{1}{9} \frac{\dot{m}}{m} R.
\end{equation}
Making use of the triangle inequality
\begin{equation}
    \vert \dot{R} \vert \leq \frac{2}{3} (\omega R) \frac{\vert \dot{\omega} \vert}{\omega^{2}} + \frac{1}{3} \frac{\vert \dot{m} \vert}{m} R.
\label{eq:triangle_inequality}
\end{equation}
Since $\vert \dot{R} \vert \equiv v_{\parallel} $ is the radial velocity of the system, while $\omega R \equiv v_{\perp}$ represents its tangential velocity, Eq. \eqref{eq:triangle_inequality} shows that the orbits are (quasi)-circular, {\it i.e.} $v_{\parallel} \ll v_{\perp}$, if $\vert \dot{\omega} \vert \ll \omega^{2}$ \textit{and} $\abs{\dot{m}} \ll m\omega$. This last condition exactly corresponds to the case $n=1$ of Eq. \eqref{eq:constraint_general}. Under those two conditions, the approximation of a circular orbit with slowly varying radius is still applicable, even in the case of a variable mass system.

In conclusion, the \textit{slowly varying mass condition}, encapsulated by the conditions \eqref{eq:constraint_general}, which enables to simplify the radiated GW power \eqref{eq:Pgw_full} to Eq. \eqref{eq:Pgw}, is linked to the condition of (quasi-)circularity of the orbital trajectories during the inspiral phase.

\medskip

To conclude this section, one should emphasize that the generic results of section \ref{sec:sec5}, although perfectly correct from a mathematical point of view, might suffer some lack of physical relevance in certain cases. Indeed, for mass evolutions presenting a stiff variation, the derivatives $\dot{m}, \ddot{m}, \dddot{m}$ cannot be legitimately neglected in the radiated power $P_{\mathrm{gw}}$ and Eq. \eqref{eq:Pgw_full} should be used instead, thus providing supplementary terms to the differential equation on $R$. Investigating the precise effect of these additional terms would constitute a work in itself, as it might as well, as explained just above, have some consequences on the form of the orbital trajectories. Let us however note that, for instance, in the context of BHs submitted to Hawking evaporation, the slowly varying mass assumption is accurately satisfied -- except for the higly fine-tuned case for which $t_{\mathrm{coal}} = t_{\mathrm{ev}}$ which might have an influence on the result of Eq. \eqref{eq:PBH_ISCO_max_freq}.

\section{\label{sec:sec7}Case of a high hierarchy of mass}
Henceforth, we have examined the case of a binary system constituted by two identical masses $m(t)$, following the same rate of variation. We now turn to the case where one of the bodies forming the binary has constant mass $M$ whereas the other body has a varying mass $m(t)$, assuming that the masses obey the hierarchy $\forall t, M \gg m(t)$. In this case, the orbital angular momentum reads 
\begin{equation}
    J_{\mathrm{orb}} \approx m \sqrt{GMR},
\label{eq:Jorb_big_mass}
\end{equation}
and its conservation leads to
\begin{equation}
    \frac{\dot{R}}{R} = -2 \frac{\dot{m}}{m}.
\end{equation}
As Kepler's third law involves the \textit{sum} of the masses of the two bodies, it now simply reads $\omega^{2} \approx GM/R^{3}$. Furthermore, the orbital energy is 
\begin{equation}
    E_{\mathrm{orbit}} = -\frac{GmM}{2R},
\label{eq:Eorbit_big_mass}
\end{equation}
so that the power radiated by the mass loss is
\begin{equation}
    P_{\mathrm{ml}} = \frac{3}{2} \frac{G\dot{m}M}{R}.
\end{equation}
One must also be careful about the expression of $P_{\mathrm{gw}}$ in this case. We established in the previous analysis that, in full generality, one must take into account the variable mass through the computations of the derivatives of the second mass moments. It is still the case for a high hierarchy  because the second mass moments $M_{ij}$ depend on the reduced mass $\mu \approx m$ for $M \gg m$.

Since Kepler's third law shows that $\omega^{2} \propto M$, the circularity condition of the orbit is satisfied only by demanding $\abs{\dot{\omega}} \ll \omega^{2}$, as in the standard case of constant masses. Consequently, if one still wants to get rid of the higher derivatives terms of $m$ in the expression of $P_{\mathrm{gw}}$ (\textit{i.e.} $\dot{m}, \ddot{m}$ and $\dddot{m}$), it boils down to an additional assumption put by hand, which has -- contrary to the case of identical evaporating masses -- no link whatsoever with the circularity of the orbit. Under this hypothesis, one can use 
\begin{equation}
    P_{\mathrm{gw}} = \frac{32}{5} \frac{c^{5}}{G} \left(\frac{GM_{c} \omega}{c^{3}}\right)^{10/3} \approx \frac{32}{5} \frac{G^4}{c^5} \frac{M^{3}m^{2}}{R^{5}}.
\end{equation}
The generic differential equation satisfied by the orbital separation ({\it i.e.} the counterpart of Eq. \eqref{eq:ED_general}) is consequently 
\begin{equation}
    \dot{R} = - \frac{64}{5} \frac{G^3}{c^5} \frac{M^{2}m}{R^{3}} -2 \frac{\dot{m}}{m}R,
\end{equation}
which is again a Bernoulli differential equation which can be analytically integrated given an explicit expression of the mass. If one takes the same mass evolutions as the one given in Eq. \eqref{eq:generic_mass}, then Eqs. \eqref{eq:separation_generic} and \eqref{eq:coal_generic} are still valid up to the following systematic substitution (in the exponents as well as in some prefactors, the various signs being unchanged)
\begin{equation}
    \frac{3}{k} \longrightarrow \frac{2}{k} \qq{and} \frac{k}{15 \pm k} \longrightarrow \frac{k}{9\pm k}.
\end{equation}
In addition, the typical time $t_{\mathrm{cc}}$ now reads
\begin{equation}
    t_{\mathrm{cc}} \equiv \frac{5}{256} \frac{c^{5} R_{0}^{4}}{G^{3} m^{2}_{\mathrm{tot}} \mu_{0}} \approx \frac{5}{256} \frac{c^{5} R_{0}^{4}}{G^{3} M^{2}m_{0}}.
\label{eq:new_tcc}
\end{equation}

\bigskip

One can also investigate the converse situation for which it is the biggest mass that varies in time {\it i.e.} $\forall t, M(t) \gg m$ and $m$ is constant. Let us note that for Hawking radiation for instance, this situation is physically irrelevant since bigger BHs are less sensitive to the effect. For completeness, however, we shall treat all the cases. The conservation of the orbital angular momentum, using Eq. \eqref{eq:Jorb_big_mass}, leads this time to
\begin{equation}
    \frac{\dot{R}}{R} = - \frac{\dot{M}}{M}.
\end{equation}
In the expression of the orbital energy, and thus in the power radiated by the variation of mass, $m$ and $M$ play symmetric roles, so Eq. \eqref{eq:Eorbit_big_mass} is unchanged and we have
\begin{equation}
    P_{\mathrm{ml}} = \frac{G \dot{M}m}{R}.
\end{equation}
The conservation of the energy leads to the following (Bernoulli) differential equation for the orbital separation
\begin{equation}
    \dot{R} = - \frac{64}{5} \frac{G^{3}}{c^{5}} \frac{M^{2}m}{R^{3}} - \frac{\dot{M}}{M}R.
\end{equation}
 If one takes the same mass evolutions as the ones given by Eq. \eqref{eq:generic_mass}, then Eqs. \eqref{eq:separation_generic} and \eqref{eq:coal_generic} are still valid up to the following systematic substitution (in the exponents as well as in some prefactors, the various signs being unchanged)
\begin{equation}
    \frac{3}{k} \longrightarrow \frac{1}{k} \qq{and} \frac{k}{15 \pm k} \longrightarrow \frac{k}{6\pm k}
\end{equation}
and the typical time $t_{\mathrm{cc}}$ is still given by Eq. \eqref{eq:new_tcc}.

\section{Conclusion}
Somehow surprisingly, even at lowest order, the dynamics of binary systems with varying mass is highly non-trivial when the emission of gravitational waves is taken into account.

In this work, we have investigated in details the evolution of binaries composed of evaporating primordial black holes with equal masses. We have shown that, depending on the initial conditions, there exist three different regimes. Both the orbital separation and the frequency of emitted gravitational waves are studied under the assumption of circularity. The critical case corresponding to an evaporation time exactly equal to the coalescence time was also considered.

Then, we focused on the case of a Bondi accretion of phantom dark energy. We took this opportunity to correct a mistake made in the literature and to settle a controversy about possible observations of this effect. This required to consider the strain, the frequency and the frequency drift of emitted gravitational waves.

Building on those results, a general study of all possible (power law) cases, including both mass losses and mass gains, was proposed. We gave full analytical solutions and, interestingly, suggested a taxonomy based only on variables easy to determine {\it a priori} -- that is the typical variation time-scale and the time the coalescence would take with constant masses (not the actual coalescence time). We have shown that quite a few different situations appear, depending on initial conditions, on the sign of the mass variation, and on the regular or ``explosive" nature of this variation. The existence of new non-monotonic regimes is underlined.

Finally, we have carefully stated the domain of validity of the different assumptions performed throughout the work and considered also the case of a high mass hierarchy between the inspiralling (or outspiralling) black holes.

In the future, it would be welcome to extend this work to the case of eccentric orbits and to investigate deeper the possible phenomenological consequences. 

\bibliographystyle{apsrev4-1}
\bibliography{apssamp}% Produces the bibliography via BibTeX.

\providecommand{\noopsort}[1]{}\providecommand{\singleletter}[1]{#1}%
\begin{thebibliography}{38}%
\makeatletter
\providecommand \@ifxundefined [1]{%
 \@ifx{#1\undefined}
}%
\providecommand \@ifnum [1]{%
 \ifnum #1\expandafter \@firstoftwo
 \else \expandafter \@secondoftwo
 \fi
}%
\providecommand \@ifx [1]{%
 \ifx #1\expandafter \@firstoftwo
 \else \expandafter \@secondoftwo
 \fi
}%
\providecommand \natexlab [1]{#1}%
\providecommand \enquote  [1]{``#1''}%
\providecommand \bibnamefont  [1]{#1}%
\providecommand \bibfnamefont [1]{#1}%
\providecommand \citenamefont [1]{#1}%
\providecommand \href@noop [0]{\@secondoftwo}%
\providecommand \href [0]{\begingroup \@sanitize@url \@href}%
\providecommand \@href[1]{\@@startlink{#1}\@@href}%
\providecommand \@@href[1]{\endgroup#1\@@endlink}%
\providecommand \@sanitize@url [0]{\catcode `\\12\catcode `\$12\catcode
  `\&12\catcode `\#12\catcode `\^12\catcode `\_12\catcode `\%12\relax}%
\providecommand \@@startlink[1]{}%
\providecommand \@@endlink[0]{}%
\providecommand \url  [0]{\begingroup\@sanitize@url \@url }%
\providecommand \@url [1]{\endgroup\@href {#1}{\urlprefix }}%
\providecommand \urlprefix  [0]{URL }%
\providecommand \Eprint [0]{\href }%
\providecommand \doibase [0]{http://dx.doi.org/}%
\providecommand \selectlanguage [0]{\@gobble}%
\providecommand \bibinfo  [0]{\@secondoftwo}%
\providecommand \bibfield  [0]{\@secondoftwo}%
\providecommand \translation [1]{[#1]}%
\providecommand \BibitemOpen [0]{}%
\providecommand \bibitemStop [0]{}%
\providecommand \bibitemNoStop [0]{.\EOS\space}%
\providecommand \EOS [0]{\spacefactor3000\relax}%
\providecommand \BibitemShut  [1]{\csname bibitem#1\endcsname}%
\let\auto@bib@innerbib\@empty
%</preamble>
\bibitem [{\citenamefont {Duerr}(2019)}]{Duerr:2019fsv}%
  \BibitemOpen
  \bibfield  {author} {\bibinfo {author} {\bibfnamefont {P.~M.}\ \bibnamefont
  {Duerr}},\ }\href {\doibase 10.1016/j.shpsb.2018.08.005} {\bibfield
  {journal} {\bibinfo  {journal} {Stud. Hist. Phil. Sci. B}\ }\textbf {\bibinfo
  {volume} {65}},\ \bibinfo {pages} {25} (\bibinfo {year} {2019})}\BibitemShut
  {NoStop}%
\bibitem [{\citenamefont {Gomes}\ and\ \citenamefont
  {Rovelli}(2023)}]{Gomes:2023xda}%
  \BibitemOpen
  \bibfield  {author} {\bibinfo {author} {\bibfnamefont {H.}~\bibnamefont
  {Gomes}}\ and\ \bibinfo {author} {\bibfnamefont {C.}~\bibnamefont
  {Rovelli}},\ }\href@noop {} {\  (\bibinfo {year} {2023})},\ \Eprint
  {http://arxiv.org/abs/2303.14064} {arXiv:2303.14064 [physics.hist-ph]}
  \BibitemShut {NoStop}%
\bibitem [{\citenamefont {Abbott}\ \emph
  {et~al.}(2021{\natexlab{a}})\citenamefont {Abbott} \emph
  {et~al.}}]{LIGOScientific:2020ibl}%
  \BibitemOpen
  \bibfield  {author} {\bibinfo {author} {\bibfnamefont {R.}~\bibnamefont
  {Abbott}} \emph {et~al.} (\bibinfo {collaboration} {LIGO Scientific,
  Virgo}),\ }\href {\doibase 10.1103/PhysRevX.11.021053} {\bibfield  {journal}
  {\bibinfo  {journal} {Phys. Rev. X}\ }\textbf {\bibinfo {volume} {11}},\
  \bibinfo {pages} {021053} (\bibinfo {year} {2021}{\natexlab{a}})},\ \Eprint
  {http://arxiv.org/abs/2010.14527} {arXiv:2010.14527 [gr-qc]} \BibitemShut
  {NoStop}%
\bibitem [{\citenamefont {Abbott}\ \emph
  {et~al.}(2021{\natexlab{b}})\citenamefont {Abbott} \emph
  {et~al.}}]{LIGOScientific:2021djp}%
  \BibitemOpen
  \bibfield  {author} {\bibinfo {author} {\bibfnamefont {R.}~\bibnamefont
  {Abbott}} \emph {et~al.} (\bibinfo {collaboration} {LIGO Scientific, VIRGO,
  KAGRA}),\ }\href@noop {} {\  (\bibinfo {year} {2021}{\natexlab{b}})},\
  \Eprint {http://arxiv.org/abs/2111.03606} {arXiv:2111.03606 [gr-qc]}
  \BibitemShut {NoStop}%
\bibitem [{\citenamefont {Holgado}\ and\ \citenamefont
  {Ricker}(2019)}]{Holgado:2019ndl}%
  \BibitemOpen
  \bibfield  {author} {\bibinfo {author} {\bibfnamefont {M.}~\bibnamefont
  {Holgado}}\ and\ \bibinfo {author} {\bibfnamefont {P.}~\bibnamefont
  {Ricker}},\ }\href {\doibase 10.3847/1538-4357/ab3293} {\bibfield  {journal}
  {\bibinfo  {journal} {The Astrophysical Journal}\ }\textbf {\bibinfo {volume}
  {882}},\ \bibinfo {pages} {39} (\bibinfo {year} {2019})}\BibitemShut
  {NoStop}%
\bibitem [{\citenamefont {Barausse}\ \emph {et~al.}(2014)\citenamefont
  {Barausse}, \citenamefont {Cardoso},\ and\ \citenamefont
  {Pani}}]{Barausse:2014tra}%
  \BibitemOpen
  \bibfield  {author} {\bibinfo {author} {\bibfnamefont {E.}~\bibnamefont
  {Barausse}}, \bibinfo {author} {\bibfnamefont {V.}~\bibnamefont {Cardoso}}, \
  and\ \bibinfo {author} {\bibfnamefont {P.}~\bibnamefont {Pani}},\ }\href
  {\doibase 10.1103/PhysRevD.89.104059} {\bibfield  {journal} {\bibinfo
  {journal} {Phys. Rev. D}\ }\textbf {\bibinfo {volume} {89}},\ \bibinfo
  {pages} {104059} (\bibinfo {year} {2014})},\ \Eprint
  {http://arxiv.org/abs/1404.7149} {arXiv:1404.7149 [gr-qc]} \BibitemShut
  {NoStop}%
\bibitem [{\citenamefont {Mersini-Houghton}\ and\ \citenamefont
  {Kelleher}(2009)}]{Mersini-Houghton:2008cul}%
  \BibitemOpen
  \bibfield  {author} {\bibinfo {author} {\bibfnamefont {L.}~\bibnamefont
  {Mersini-Houghton}}\ and\ \bibinfo {author} {\bibfnamefont {A.}~\bibnamefont
  {Kelleher}},\ }\href {\doibase
  https://doi.org/10.1016/j.nuclphysbps.2009.07.091} {\bibfield  {journal}
  {\bibinfo  {journal} {Nuclear Physics B - Proceedings Supplements}\ }\textbf
  {\bibinfo {volume} {194}},\ \bibinfo {pages} {272} (\bibinfo {year}
  {2009})},\ \bibinfo {note} {new Horizons for Modern Cosmology}\BibitemShut
  {NoStop}%
\bibitem [{\citenamefont {Enander}\ and\ \citenamefont
  {Mörtsell}(2010)}]{Enander_2010}%
  \BibitemOpen
  \bibfield  {author} {\bibinfo {author} {\bibfnamefont {J.}~\bibnamefont
  {Enander}}\ and\ \bibinfo {author} {\bibfnamefont {E.}~\bibnamefont
  {Mörtsell}},\ }\href {\doibase 10.1016/j.physletb.2009.11.057} {\bibfield
  {journal} {\bibinfo  {journal} {Physics Letters B}\ }\textbf {\bibinfo
  {volume} {683}},\ \bibinfo {pages} {7} (\bibinfo {year} {2010})}\BibitemShut
  {NoStop}%
\bibitem [{\citenamefont {O'Neill}\ \emph {et~al.}(2009)\citenamefont
  {O'Neill}, \citenamefont {Miller}, \citenamefont {Bogdanovic}, \citenamefont
  {Reynolds},\ and\ \citenamefont {Schnittman}}]{ONeill:2008sat}%
  \BibitemOpen
  \bibfield  {author} {\bibinfo {author} {\bibfnamefont {S.~M.}\ \bibnamefont
  {O'Neill}}, \bibinfo {author} {\bibfnamefont {M.~C.}\ \bibnamefont {Miller}},
  \bibinfo {author} {\bibfnamefont {T.}~\bibnamefont {Bogdanovic}}, \bibinfo
  {author} {\bibfnamefont {C.~S.}\ \bibnamefont {Reynolds}}, \ and\ \bibinfo
  {author} {\bibfnamefont {J.}~\bibnamefont {Schnittman}},\ }\href {\doibase
  10.1088/0004-637X/700/1/859} {\bibfield  {journal} {\bibinfo  {journal}
  {Astrophys. J.}\ }\textbf {\bibinfo {volume} {700}},\ \bibinfo {pages} {859}
  (\bibinfo {year} {2009})},\ \Eprint {http://arxiv.org/abs/0812.4874}
  {arXiv:0812.4874 [astro-ph]} \BibitemShut {NoStop}%
\bibitem [{\citenamefont {Macedo}\ \emph {et~al.}(2013)\citenamefont {Macedo},
  \citenamefont {Pani}, \citenamefont {Cardoso},\ and\ \citenamefont
  {Crispino}}]{Macedo:2013qea}%
  \BibitemOpen
  \bibfield  {author} {\bibinfo {author} {\bibfnamefont {C.~F.~B.}\
  \bibnamefont {Macedo}}, \bibinfo {author} {\bibfnamefont {P.}~\bibnamefont
  {Pani}}, \bibinfo {author} {\bibfnamefont {V.}~\bibnamefont {Cardoso}}, \
  and\ \bibinfo {author} {\bibfnamefont {L.~C.~B.}\ \bibnamefont {Crispino}},\
  }\href {\doibase 10.1088/0004-637X/774/1/48} {\bibfield  {journal} {\bibinfo
  {journal} {Astrophys. J.}\ }\textbf {\bibinfo {volume} {774}},\ \bibinfo
  {pages} {48} (\bibinfo {year} {2013})},\ \Eprint
  {http://arxiv.org/abs/1302.2646} {arXiv:1302.2646 [gr-qc]} \BibitemShut
  {NoStop}%
\bibitem [{\citenamefont {Sarkar}\ \emph {et~al.}(2019)\citenamefont {Sarkar},
  \citenamefont {Rajesh~Nayak},\ and\ \citenamefont
  {Majumdar}}]{Sarkar:2019tdf}%
  \BibitemOpen
  \bibfield  {author} {\bibinfo {author} {\bibfnamefont {A.}~\bibnamefont
  {Sarkar}}, \bibinfo {author} {\bibfnamefont {K.}~\bibnamefont
  {Rajesh~Nayak}}, \ and\ \bibinfo {author} {\bibfnamefont {A.~S.}\
  \bibnamefont {Majumdar}},\ }\href {\doibase 10.1103/PhysRevD.100.103514}
  {\bibfield  {journal} {\bibinfo  {journal} {Phys. Rev. D}\ }\textbf {\bibinfo
  {volume} {100}},\ \bibinfo {pages} {103514} (\bibinfo {year} {2019})},\
  \Eprint {http://arxiv.org/abs/1904.13261} {arXiv:1904.13261 [astro-ph.CO]}
  \BibitemShut {NoStop}%
\bibitem [{\citenamefont {Sarkar}\ \emph {et~al.}(2023)\citenamefont {Sarkar},
  \citenamefont {Ali}, \citenamefont {Nayak},\ and\ \citenamefont
  {Majumdar}}]{Sarkar:2022jtn}%
  \BibitemOpen
  \bibfield  {author} {\bibinfo {author} {\bibfnamefont {A.}~\bibnamefont
  {Sarkar}}, \bibinfo {author} {\bibfnamefont {A.}~\bibnamefont {Ali}},
  \bibinfo {author} {\bibfnamefont {K.~R.}\ \bibnamefont {Nayak}}, \ and\
  \bibinfo {author} {\bibfnamefont {A.~S.}\ \bibnamefont {Majumdar}},\ }\href
  {\doibase 10.1103/PhysRevD.107.084038} {\bibfield  {journal} {\bibinfo
  {journal} {Phys. Rev. D}\ }\textbf {\bibinfo {volume} {107}},\ \bibinfo
  {pages} {084038} (\bibinfo {year} {2023})},\ \Eprint
  {http://arxiv.org/abs/2210.12502} {arXiv:2210.12502 [gr-qc]} \BibitemShut
  {NoStop}%
\bibitem [{\citenamefont {Yagi}\ \emph {et~al.}(2011)\citenamefont {Yagi},
  \citenamefont {Tanahashi},\ and\ \citenamefont {Tanaka}}]{Yagi:2011yu}%
  \BibitemOpen
  \bibfield  {author} {\bibinfo {author} {\bibfnamefont {K.}~\bibnamefont
  {Yagi}}, \bibinfo {author} {\bibfnamefont {N.}~\bibnamefont {Tanahashi}}, \
  and\ \bibinfo {author} {\bibfnamefont {T.}~\bibnamefont {Tanaka}},\ }\href
  {\doibase 10.1103/PhysRevD.83.084036} {\bibfield  {journal} {\bibinfo
  {journal} {Phys. Rev. D}\ }\textbf {\bibinfo {volume} {83}},\ \bibinfo
  {pages} {084036} (\bibinfo {year} {2011})},\ \Eprint
  {http://arxiv.org/abs/1101.4997} {arXiv:1101.4997 [gr-qc]} \BibitemShut
  {NoStop}%
\bibitem [{\citenamefont {Simonetti}\ \emph {et~al.}(2011)\citenamefont
  {Simonetti}, \citenamefont {Kavic}, \citenamefont {Minic}, \citenamefont
  {Surani},\ and\ \citenamefont {Vijayan}}]{Simonetti:2010mk}%
  \BibitemOpen
  \bibfield  {author} {\bibinfo {author} {\bibfnamefont {J.~H.}\ \bibnamefont
  {Simonetti}}, \bibinfo {author} {\bibfnamefont {M.}~\bibnamefont {Kavic}},
  \bibinfo {author} {\bibfnamefont {D.}~\bibnamefont {Minic}}, \bibinfo
  {author} {\bibfnamefont {U.}~\bibnamefont {Surani}}, \ and\ \bibinfo {author}
  {\bibfnamefont {V.}~\bibnamefont {Vijayan}},\ }\href {\doibase
  10.1088/2041-8205/737/2/L28} {\bibfield  {journal} {\bibinfo  {journal}
  {Astrophys. J. Lett.}\ }\textbf {\bibinfo {volume} {737}},\ \bibinfo {pages}
  {L28} (\bibinfo {year} {2011})},\ \Eprint {http://arxiv.org/abs/1010.5245}
  {arXiv:1010.5245 [astro-ph.HE]} \BibitemShut {NoStop}%
\bibitem [{\citenamefont {Yuan}\ \emph {et~al.}(2021)\citenamefont {Yuan},
  \citenamefont {Brito},\ and\ \citenamefont {Cardoso}}]{Yuan:2021xdi}%
  \BibitemOpen
  \bibfield  {author} {\bibinfo {author} {\bibfnamefont {C.}~\bibnamefont
  {Yuan}}, \bibinfo {author} {\bibfnamefont {R.}~\bibnamefont {Brito}}, \ and\
  \bibinfo {author} {\bibfnamefont {V.}~\bibnamefont {Cardoso}},\ }\href
  {\doibase 10.1103/PhysRevD.104.124024} {\bibfield  {journal} {\bibinfo
  {journal} {Phys. Rev. D}\ }\textbf {\bibinfo {volume} {104}},\ \bibinfo
  {pages} {124024} (\bibinfo {year} {2021})},\ \Eprint
  {http://arxiv.org/abs/2107.14244} {arXiv:2107.14244 [gr-qc]} \BibitemShut
  {NoStop}%
\bibitem [{\citenamefont {Maggiore}(2007)}]{Maggiore:2007ulw}%
  \BibitemOpen
  \bibfield  {author} {\bibinfo {author} {\bibfnamefont {M.}~\bibnamefont
  {Maggiore}},\ }\href {\doibase 10.1093/acprof:oso/9780198570745.001.0001}
  {\emph {\bibinfo {title} {{Gravitational Waves. Vol. 1: Theory and
  Experiments}}}}\ (\bibinfo  {publisher} {Oxford University Press},\ \bibinfo
  {year} {2007})\BibitemShut {NoStop}%
\bibitem [{\citenamefont {Aggarwal}\ \emph {et~al.}(2021)\citenamefont
  {Aggarwal} \emph {et~al.}}]{Aggarwal:2020olq}%
  \BibitemOpen
  \bibfield  {author} {\bibinfo {author} {\bibfnamefont {N.}~\bibnamefont
  {Aggarwal}} \emph {et~al.},\ }\href {\doibase 10.1007/s41114-021-00032-5}
  {\bibfield  {journal} {\bibinfo  {journal} {Living Rev. Rel.}\ }\textbf
  {\bibinfo {volume} {24}},\ \bibinfo {pages} {4} (\bibinfo {year} {2021})},\
  \Eprint {http://arxiv.org/abs/2011.12414} {arXiv:2011.12414 [gr-qc]}
  \BibitemShut {NoStop}%
\bibitem [{\citenamefont {Babichev}\ \emph {et~al.}(2004)\citenamefont
  {Babichev}, \citenamefont {Dokuchaev},\ and\ \citenamefont
  {Eroshenko}}]{Babichev:2004yx}%
  \BibitemOpen
  \bibfield  {author} {\bibinfo {author} {\bibfnamefont {E.}~\bibnamefont
  {Babichev}}, \bibinfo {author} {\bibfnamefont {V.}~\bibnamefont {Dokuchaev}},
  \ and\ \bibinfo {author} {\bibfnamefont {Y.}~\bibnamefont {Eroshenko}},\
  }\href {\doibase 10.1103/PhysRevLett.93.021102} {\bibfield  {journal}
  {\bibinfo  {journal} {Phys. Rev. Lett.}\ }\textbf {\bibinfo {volume} {93}},\
  \bibinfo {pages} {021102} (\bibinfo {year} {2004})},\ \Eprint
  {http://arxiv.org/abs/gr-qc/0402089} {arXiv:gr-qc/0402089} \BibitemShut
  {NoStop}%
\bibitem [{\citenamefont {Bondi}(1952)}]{Bondi1952}%
  \BibitemOpen
  \bibfield  {author} {\bibinfo {author} {\bibfnamefont {H.}~\bibnamefont
  {Bondi}},\ }\href {\doibase 10.1093/mnras/112.2.195} {\bibfield  {journal}
  {\bibinfo  {journal} {Monthly Notices of the Royal Astronomical Society}\
  }\textbf {\bibinfo {volume} {112}},\ \bibinfo {pages} {195} (\bibinfo {year}
  {1952})},\ \Eprint
  {http://arxiv.org/abs/https://academic.oup.com/mnras/article-pdf/112/2/195/9073555/mnras112-0195.pdf}
  {https://academic.oup.com/mnras/article-pdf/112/2/195/9073555/mnras112-0195.pdf}
  \BibitemShut {NoStop}%
\bibitem [{\citenamefont {Babichev}\ \emph {et~al.}(2005)\citenamefont
  {Babichev}, \citenamefont {Dokuchaev},\ and\ \citenamefont
  {Eroshenko}}]{Babichev:2005py}%
  \BibitemOpen
  \bibfield  {author} {\bibinfo {author} {\bibfnamefont {E.}~\bibnamefont
  {Babichev}}, \bibinfo {author} {\bibfnamefont {V.}~\bibnamefont {Dokuchaev}},
  \ and\ \bibinfo {author} {\bibfnamefont {Y.}~\bibnamefont {Eroshenko}},\
  }\href {\doibase 10.1134/1.1901765} {\bibfield  {journal} {\bibinfo
  {journal} {J. Exp. Theor. Phys.}\ }\textbf {\bibinfo {volume} {100}},\
  \bibinfo {pages} {528} (\bibinfo {year} {2005})},\ \Eprint
  {http://arxiv.org/abs/astro-ph/0505618} {arXiv:astro-ph/0505618} \BibitemShut
  {NoStop}%
\bibitem [{\citenamefont {Carroll}\ \emph {et~al.}(2003)\citenamefont
  {Carroll}, \citenamefont {Hoffman},\ and\ \citenamefont
  {Trodden}}]{Carroll:2003st}%
  \BibitemOpen
  \bibfield  {author} {\bibinfo {author} {\bibfnamefont {S.~M.}\ \bibnamefont
  {Carroll}}, \bibinfo {author} {\bibfnamefont {M.}~\bibnamefont {Hoffman}}, \
  and\ \bibinfo {author} {\bibfnamefont {M.}~\bibnamefont {Trodden}},\ }\href
  {\doibase 10.1103/PhysRevD.68.023509} {\bibfield  {journal} {\bibinfo
  {journal} {Phys. Rev. D}\ }\textbf {\bibinfo {volume} {68}},\ \bibinfo
  {pages} {023509} (\bibinfo {year} {2003})},\ \Eprint
  {http://arxiv.org/abs/astro-ph/0301273} {arXiv:astro-ph/0301273} \BibitemShut
  {NoStop}%
\bibitem [{\citenamefont {He}\ \emph {et~al.}(2009)\citenamefont {He},
  \citenamefont {Wang}, \citenamefont {Wu},\ and\ \citenamefont
  {Lin}}]{He:2009jd}%
  \BibitemOpen
  \bibfield  {author} {\bibinfo {author} {\bibfnamefont {X.}~\bibnamefont
  {He}}, \bibinfo {author} {\bibfnamefont {B.}~\bibnamefont {Wang}}, \bibinfo
  {author} {\bibfnamefont {S.-F.}\ \bibnamefont {Wu}}, \ and\ \bibinfo {author}
  {\bibfnamefont {C.-Y.}\ \bibnamefont {Lin}},\ }\href {\doibase
  10.1016/j.physletb.2009.02.002} {\bibfield  {journal} {\bibinfo  {journal}
  {Phys. Lett. B}\ }\textbf {\bibinfo {volume} {673}},\ \bibinfo {pages} {156}
  (\bibinfo {year} {2009})},\ \Eprint {http://arxiv.org/abs/0901.0034}
  {arXiv:0901.0034 [gr-qc]} \BibitemShut {NoStop}%
\bibitem [{\citenamefont {Hadjidemetriou}(1966)}]{HADJIDEMETRIOU196634}%
  \BibitemOpen
  \bibfield  {author} {\bibinfo {author} {\bibfnamefont {J.~D.}\ \bibnamefont
  {Hadjidemetriou}},\ }\href {\doibase
  https://doi.org/10.1016/0019-1035(66)90006-6} {\bibfield  {journal} {\bibinfo
   {journal} {Icarus}\ }\textbf {\bibinfo {volume} {5}},\ \bibinfo {pages} {34}
  (\bibinfo {year} {1966})}\BibitemShut {NoStop}%
\bibitem [{\citenamefont {{Verhulst}}(1975)}]{Verhulst1975}%
  \BibitemOpen
  \bibfield  {author} {\bibinfo {author} {\bibfnamefont {F.}~\bibnamefont
  {{Verhulst}}},\ }\href {\doibase 10.1007/BF01228739} {\bibfield  {journal}
  {\bibinfo  {journal} {Celestial Mechanics}\ }\textbf {\bibinfo {volume}
  {11}},\ \bibinfo {pages} {95} (\bibinfo {year} {1975})}\BibitemShut {NoStop}%
\bibitem [{\citenamefont {McWilliams}(2010)}]{McWilliams:2009ym}%
  \BibitemOpen
  \bibfield  {author} {\bibinfo {author} {\bibfnamefont {S.~T.}\ \bibnamefont
  {McWilliams}},\ }\href {\doibase 10.1103/PhysRevLett.104.141601} {\bibfield
  {journal} {\bibinfo  {journal} {Phys. Rev. Lett.}\ }\textbf {\bibinfo
  {volume} {104}},\ \bibinfo {pages} {141601} (\bibinfo {year} {2010})},\
  \Eprint {http://arxiv.org/abs/0912.4744} {arXiv:0912.4744 [gr-qc]}
  \BibitemShut {NoStop}%
\bibitem [{\citenamefont {Carr}\ \emph {et~al.}(2021)\citenamefont {Carr},
  \citenamefont {Kohri}, \citenamefont {Sendouda},\ and\ \citenamefont
  {Yokoyama}}]{Carr:2020gox}%
  \BibitemOpen
  \bibfield  {author} {\bibinfo {author} {\bibfnamefont {B.}~\bibnamefont
  {Carr}}, \bibinfo {author} {\bibfnamefont {K.}~\bibnamefont {Kohri}},
  \bibinfo {author} {\bibfnamefont {Y.}~\bibnamefont {Sendouda}}, \ and\
  \bibinfo {author} {\bibfnamefont {J.}~\bibnamefont {Yokoyama}},\ }\href
  {\doibase 10.1088/1361-6633/ac1e31} {\bibfield  {journal} {\bibinfo
  {journal} {Rept. Prog. Phys.}\ }\textbf {\bibinfo {volume} {84}},\ \bibinfo
  {pages} {116902} (\bibinfo {year} {2021})},\ \Eprint
  {http://arxiv.org/abs/2002.12778} {arXiv:2002.12778 [astro-ph.CO]}
  \BibitemShut {NoStop}%
\bibitem [{\citenamefont {MacGibbon}(1991)}]{MacGibbon_PBH}%
  \BibitemOpen
  \bibfield  {author} {\bibinfo {author} {\bibfnamefont {J.~H.}\ \bibnamefont
  {MacGibbon}},\ }\href {\doibase 10.1103/PhysRevD.44.376} {\bibfield
  {journal} {\bibinfo  {journal} {Phys. Rev. D}\ }\textbf {\bibinfo {volume}
  {44}},\ \bibinfo {pages} {376} (\bibinfo {year} {1991})}\BibitemShut
  {NoStop}%
\bibitem [{\citenamefont {Halzen}\ \emph {et~al.}(1991)\citenamefont {Halzen},
  \citenamefont {Zas}, \citenamefont {MacGibbon},\ and\ \citenamefont
  {Weekes}}]{Halzen:1991uw}%
  \BibitemOpen
  \bibfield  {author} {\bibinfo {author} {\bibfnamefont {F.}~\bibnamefont
  {Halzen}}, \bibinfo {author} {\bibfnamefont {E.}~\bibnamefont {Zas}},
  \bibinfo {author} {\bibfnamefont {J.~H.}\ \bibnamefont {MacGibbon}}, \ and\
  \bibinfo {author} {\bibfnamefont {T.~C.}\ \bibnamefont {Weekes}},\ }\href
  {\doibase 10.1038/353807a0} {\bibfield  {journal} {\bibinfo  {journal}
  {Nature}\ }\textbf {\bibinfo {volume} {353}},\ \bibinfo {pages} {807}
  (\bibinfo {year} {1991})}\BibitemShut {NoStop}%
\bibitem [{\citenamefont {Hawking}(1975)}]{Hawking:1975vcx}%
  \BibitemOpen
  \bibfield  {author} {\bibinfo {author} {\bibfnamefont {S.~W.}\ \bibnamefont
  {Hawking}},\ }\href {\doibase 10.1007/BF02345020} {\bibfield  {journal}
  {\bibinfo  {journal} {Commun. Math. Phys.}\ }\textbf {\bibinfo {volume}
  {43}},\ \bibinfo {pages} {199} (\bibinfo {year} {1975})},\ \bibinfo {note}
  {[Erratum: Commun.Math.Phys. 46, 206 (1976)]}\BibitemShut {NoStop}%
\bibitem [{\citenamefont {Birrell}\ and\ \citenamefont
  {Davies}(1984)}]{Birrell:1982ix}%
  \BibitemOpen
  \bibfield  {author} {\bibinfo {author} {\bibfnamefont {N.~D.}\ \bibnamefont
  {Birrell}}\ and\ \bibinfo {author} {\bibfnamefont {P.~C.~W.}\ \bibnamefont
  {Davies}},\ }\href {\doibase 10.1017/CBO9780511622632} {\emph {\bibinfo
  {title} {{Quantum Fields in Curved Space}}}},\ Cambridge Monographs on
  Mathematical Physics\ (\bibinfo  {publisher} {Cambridge Univ. Press},\
  \bibinfo {address} {Cambridge, UK},\ \bibinfo {year} {1984})\BibitemShut
  {NoStop}%
\bibitem [{\citenamefont {Kocsis}\ \emph {et~al.}(2018)\citenamefont {Kocsis},
  \citenamefont {Suyama}, \citenamefont {Tanaka},\ and\ \citenamefont
  {Yokoyama}}]{Kocsis:2017yty}%
  \BibitemOpen
  \bibfield  {author} {\bibinfo {author} {\bibfnamefont {B.}~\bibnamefont
  {Kocsis}}, \bibinfo {author} {\bibfnamefont {T.}~\bibnamefont {Suyama}},
  \bibinfo {author} {\bibfnamefont {T.}~\bibnamefont {Tanaka}}, \ and\ \bibinfo
  {author} {\bibfnamefont {S.}~\bibnamefont {Yokoyama}},\ }\href {\doibase
  10.3847/1538-4357/aaa7f4} {\bibfield  {journal} {\bibinfo  {journal}
  {Astrophys. J.}\ }\textbf {\bibinfo {volume} {854}},\ \bibinfo {pages} {41}
  (\bibinfo {year} {2018})},\ \Eprint {http://arxiv.org/abs/1709.09007}
  {arXiv:1709.09007 [astro-ph.CO]} \BibitemShut {NoStop}%
\bibitem [{\citenamefont {Raidal}\ \emph {et~al.}(2019)\citenamefont {Raidal},
  \citenamefont {Spethmann}, \citenamefont {Vaskonen},\ and\ \citenamefont
  {Veerm\"ae}}]{Raidal:2018bbj}%
  \BibitemOpen
  \bibfield  {author} {\bibinfo {author} {\bibfnamefont {M.}~\bibnamefont
  {Raidal}}, \bibinfo {author} {\bibfnamefont {C.}~\bibnamefont {Spethmann}},
  \bibinfo {author} {\bibfnamefont {V.}~\bibnamefont {Vaskonen}}, \ and\
  \bibinfo {author} {\bibfnamefont {H.}~\bibnamefont {Veerm\"ae}},\ }\href
  {\doibase 10.1088/1475-7516/2019/02/018} {\bibfield  {journal} {\bibinfo
  {journal} {JCAP}\ }\textbf {\bibinfo {volume} {02}},\ \bibinfo {pages} {018}
  (\bibinfo {year} {2019})},\ \Eprint {http://arxiv.org/abs/1812.01930}
  {arXiv:1812.01930 [astro-ph.CO]} \BibitemShut {NoStop}%
\bibitem [{\citenamefont {Gow}\ \emph {et~al.}(2020)\citenamefont {Gow},
  \citenamefont {Byrnes}, \citenamefont {Hall},\ and\ \citenamefont
  {Peacock}}]{Gow:2019pok}%
  \BibitemOpen
  \bibfield  {author} {\bibinfo {author} {\bibfnamefont {A.~D.}\ \bibnamefont
  {Gow}}, \bibinfo {author} {\bibfnamefont {C.~T.}\ \bibnamefont {Byrnes}},
  \bibinfo {author} {\bibfnamefont {A.}~\bibnamefont {Hall}}, \ and\ \bibinfo
  {author} {\bibfnamefont {J.~A.}\ \bibnamefont {Peacock}},\ }\href {\doibase
  10.1088/1475-7516/2020/01/031} {\bibfield  {journal} {\bibinfo  {journal}
  {JCAP}\ }\textbf {\bibinfo {volume} {01}},\ \bibinfo {pages} {031} (\bibinfo
  {year} {2020})},\ \Eprint {http://arxiv.org/abs/1911.12685} {arXiv:1911.12685
  [astro-ph.CO]} \BibitemShut {NoStop}%
\bibitem [{\citenamefont {Antelis}\ \emph {et~al.}(2018)\citenamefont
  {Antelis}, \citenamefont {Hern\'andez},\ and\ \citenamefont
  {Moreno}}]{Antelis:2018sfj}%
  \BibitemOpen
  \bibfield  {author} {\bibinfo {author} {\bibfnamefont {J.~M.}\ \bibnamefont
  {Antelis}}, \bibinfo {author} {\bibfnamefont {J.~M.}\ \bibnamefont
  {Hern\'andez}}, \ and\ \bibinfo {author} {\bibfnamefont {C.}~\bibnamefont
  {Moreno}},\ }\href {\doibase 10.1088/1742-6596/1030/1/012005} {\bibfield
  {journal} {\bibinfo  {journal} {J. Phys. Conf. Ser.}\ }\textbf {\bibinfo
  {volume} {1030}},\ \bibinfo {pages} {012005} (\bibinfo {year}
  {2018})}\BibitemShut {NoStop}%
\bibitem [{\citenamefont {Aghanim}\ \emph {et~al.}(2020)\citenamefont {Aghanim}
  \emph {et~al.}}]{Planck:2018vyg}%
  \BibitemOpen
  \bibfield  {author} {\bibinfo {author} {\bibfnamefont {N.}~\bibnamefont
  {Aghanim}} \emph {et~al.} (\bibinfo {collaboration} {Planck}),\ }\href
  {\doibase 10.1051/0004-6361/201833910} {\bibfield  {journal} {\bibinfo
  {journal} {Astron. Astrophys.}\ }\textbf {\bibinfo {volume} {641}},\ \bibinfo
  {pages} {A6} (\bibinfo {year} {2020})},\ \bibinfo {note} {[Erratum:
  Astron.Astrophys. 652, C4 (2021)]},\ \Eprint
  {http://arxiv.org/abs/1807.06209} {arXiv:1807.06209 [astro-ph.CO]}
  \BibitemShut {NoStop}%
\bibitem [{\citenamefont {Barrau}\ \emph {et~al.}(2022)\citenamefont {Barrau},
  \citenamefont {Martineau},\ and\ \citenamefont {Renevey}}]{Barrau:2022bfg}%
  \BibitemOpen
  \bibfield  {author} {\bibinfo {author} {\bibfnamefont {A.}~\bibnamefont
  {Barrau}}, \bibinfo {author} {\bibfnamefont {K.}~\bibnamefont {Martineau}}, \
  and\ \bibinfo {author} {\bibfnamefont {C.}~\bibnamefont {Renevey}},\ }\href
  {\doibase 10.1103/PhysRevD.106.023509} {\bibfield  {journal} {\bibinfo
  {journal} {Phys. Rev. D}\ }\textbf {\bibinfo {volume} {106}},\ \bibinfo
  {pages} {023509} (\bibinfo {year} {2022})},\ \Eprint
  {http://arxiv.org/abs/2203.13297} {arXiv:2203.13297 [gr-qc]} \BibitemShut
  {NoStop}%
\bibitem [{\citenamefont {Acernese}\ \emph {et~al.}(2022)\citenamefont
  {Acernese} \emph {et~al.}}]{Virgo:2022ysc}%
  \BibitemOpen
  \bibfield  {author} {\bibinfo {author} {\bibfnamefont {F.}~\bibnamefont
  {Acernese}} \emph {et~al.} (\bibinfo {collaboration} {Virgo}),\ }\href@noop
  {} {\  (\bibinfo {year} {2022})},\ \Eprint {http://arxiv.org/abs/2210.15633}
  {arXiv:2210.15633 [gr-qc]} \BibitemShut {NoStop}%
\bibitem [{\citenamefont {Barrau}\ \emph {et~al.}(2023)\citenamefont {Barrau},
  \citenamefont {Garc\'\i{}a-Bellido}, \citenamefont {Grenet},\ and\
  \citenamefont {Martineau}}]{Barrau:2023kuv}%
  \BibitemOpen
  \bibfield  {author} {\bibinfo {author} {\bibfnamefont {A.}~\bibnamefont
  {Barrau}}, \bibinfo {author} {\bibfnamefont {J.}~\bibnamefont
  {Garc\'\i{}a-Bellido}}, \bibinfo {author} {\bibfnamefont {T.}~\bibnamefont
  {Grenet}}, \ and\ \bibinfo {author} {\bibfnamefont {K.}~\bibnamefont
  {Martineau}},\ }\href@noop {} {\  (\bibinfo {year} {2023})},\ \Eprint
  {http://arxiv.org/abs/2303.06006} {arXiv:2303.06006 [gr-qc]} \BibitemShut
  {NoStop}%
\end{thebibliography}%

\end{document}